\documentclass[referee,a4paper,12pt,traditabstract]{swsc} 

\usepackage[backref]{hyperref}
\usepackage{graphicx}
\usepackage{subfigure}
\usepackage{epstopdf}
\usepackage{lineno}
\usepackage{natbib}
\usepackage{amsmath}
\begin{document}

   \title{Updated Model of the Solar Energetic Proton Environment in Space}


   \author{Piers Jiggens\inst{1}
          \and
          Daniel Heynderickx\inst{2}
          \and
          Ingmar Sandberg\inst{3}
          \and
          Pete Truscott\inst{4}
					\and
          Osku Raukunen\inst{5} 
          \and
          Rami Vainio\inst{5} 
          }

   \institute{European Space Research and Technology Centre (ESTEC), Space Environment and Effects Section
              Keperlaan 1, 2200AG Noordwijk, The Netherlands
							\email{\href{mailto:piers.jiggens@esa.int}{piers.jiggens@esa.int}}
						\and
						 DH Consultancy, Leuven, Belgium 
						\and
						 SPARC Space Applications \& Research Consultancy, Athens, Greece
						\and
             Kallisto Consultancy, Farnborough, United Kingdom
						\and
						Department of Physics and Astronomy, University of Turku, FI-20014 Turku, Finland
						}


 
  \abstract
   {
 
The Solar Accumulated and Peak Proton and Heavy Ion Radiation Environment (SAPPHIRE) model provides environment specification outputs for all aspects of the Solar Energetic Particle (SEP) environment. The model is based upon a thoroughly cleaned and carefully processed data set. Herein the evolution of the solar proton model is discussed with comparisons to other models and data. This paper discusses the construction of the underlying data set, the modelling methodology, optimisation of fitted flux distributions and extrapolation of model outputs to cover a range of proton energies from 0.1 MeV to 1 GeV. The model provides outputs in terms of mission cumulative fluence, maximum event fluence and peak flux for both solar maximum and solar minimum periods. A new method for describing maximum event fluence and peak flux outputs in terms of 1-in-x-year SPEs is also described. SAPPHIRE proton model outputs are compared with previous models including CREME96, ESP-PSYCHIC and the JPL model. Low energy outputs are compared to SEP data from ACE/EPAM whilst high energy outputs are compared to a new model based on GLEs detected by Neutron Monitors (NMs).
   }        

	\titlerunning{Updated Model of the SEP Environment in Space}
	\authorrunning{Jiggens et al.}
   \maketitle


\section{Introduction}

For space missions one critical component of the design process is to know radiation levels to be applied for the qualification of instruments and components to ensure survival while operating nominally. Total accumulated dose results in the degradation and ultimate failure of electronic components \citep{Daly1996} due to ionisation or displacement damage mechanisms \citep{Feynman2000a}. There are three (naturally occurring) components of the space radiation environment: Galactic Cosmic Rays (GCRs), radiation trapped by planetary magnetic fields (in the case of the Earth these are termed the Van Allen belts) and Solar Energetic Particles (SEPs). For spacecraft orbiting the Earth ionising dose effects are often dominated by trapped particles (protons and electrons) while displacement damage (non-ionising) dose effects in Low Earth Orbit (LEO) are often dominated by trapped protons. However, for Earth-orbiting space missions at altitudes and magnetic latitudes much beyond LEO, where the effect of trapped protons is lower, SEPs can have an important influence on radiation doses and for spacecraft which do not regularly pass through the proton belt SEPs are often the dominant source of displacement damage effects in opto-electronic components (e.g. solar array degradation). In the interplanetary environment SEPs are the dominant source for all radiation dose effects. SEPs may also be an important component in terms of Single Event Effects (SEEs) in electronic components in addition to the slowly modulating background flux of GCRs which, although low in absolute flux levels, have relatively hard spectra with high abundances of heavier particles. In radiation hardened components it is only heavier ions (such as Fe) which deposit sufficient energy to produce SEEs at non-negligible rates.

SEPs are detected in interplanetary medium and consist of electrons, protons, and heavier ions up to Fe (and higher) with energies from the tens of keV to the GeV range. SEPs are characterised by large enhancements in particle fluxes many orders of magnitude above background levels. Such enhancements are termed Solar Particle Events (SPEs). SPEs result from the acceleration of particles in the solar corona either by solar flares, by interplanetary shocks driven by Coronal Mass Ejections (CMEs) or by shocks associated with Co-rotating Interaction Regions (CIRs). Accelerated particles then propagate through the heliosphere, spiralling along the Interplanetary Magnetic Field (IMF).

Previous important work in this field includes the King (SOLPRO) model \citep{King1974} which was further developed into the JPL model by \citet{Feynman1990a,Feynman1993,Feynman2002} and associated work by \citet{Jun2007}. These models focussed on the cumulative fluence of the SEP environment which is important for the specification of radiation dose levels. Due to the highly stochastic nature of SPEs model outputs are expressed as a function of probability, or confidence that the customer requires, that a specified fluence would not be exceeded. As such models require large numbers of iterations and each of the energies of interest are addressed separately. Work by \citet{Xapsos1998,Xapsos1999} on the ESP models initially focussed on modelling the peak fluxes and worst case SPE fluences with analytical expressions to derive quantities for any mission duration and probability. These can be useful in terms of proton-induced upset rates and sensor interference. ESP was expanded to also include models of the cumulative fluence environment extrapolated from the distribution of yearly solar proton fluences and was later re-named PSYCHIC \citep{Xapsos2000,Xapsos2004,Xapsos2007}. The development of the Moscow State University (MSU) model principally by \citet{Nymmik1999,Nymmik2007,Nymmik2011} is based on a single reference energy of SEPs and uses spectral forms fit to data to derive outputs at other energies. Whilst the JPL and ESP models treat the solar cycle in two distinct phases (active and quiet), the MSU model connects the SPE frequency to the Wolf sunspot number. The PSYCHIC-ESP model is the present standard for long-term SEP fluences as specified in the relevant document from the initiative of the European Cooperation for Space Standardization (ECSS) \citep{ECSS1004C}.

Distributions applied to SEP fluxes include a lognormal distribution (JPL) and variants of power laws as used in the ESP and MSU models which were earlier applied to SPEs by \citet{Gabriel1996}. One challenge for the modelling of the SEP environment is to quantify the deviation of fluxes from a pure power law. This tail of the distribution is important because it dominates both the peak and cumulative environment calculations for standard prediction periods at the high confidence levels required for space missions due to the wide spread of SEP flux magnitudes between events (many orders of magnitude depending on energy). Given that space-based SEP data is available over only the past 5 solar cycles whilst spacecraft designers require a high confidence that specified levels will not be exceeded in a mission lifetime the role of statistics in deriving models is of critical importance. In this context is is also important to note the the spectra applied for SEE calculations, including those specified in the relevant ECSS standard \citep{ECSS1004C}, is most frequently taken from CREME96 \citep{Tylka1997a}. CREME96 takes the worst case to be the single observed SPE from October 1989 with a justification that models would extrapolate a higher value which when put into the context of the margins in radiation hardness assurance processes would lead to excessive conservatism.

The Solar Accumulated and Peak Proton and Heavy Ion Radiation Environment (SAPPHIRE) model aims to specify the SEP environment covering all associated ions, energies and timescales. The SAPPHIRE model has been developed in the context of ESA activities to develop the Solar Energetic Particle Environment Modelling (SEPEM) system \citep{Crosby2015}. The system (\url{http://sepem.eu}) allows users to browse raw and processed SEP data and to develop models based on these data including the effects of physical and magnetic shielding as well as the impact of the distance to the Sun for interplanetary missions.

This paper focusses on the derivation of the solar proton element of SAPPHIRE which can be seen as a significant update of earlier work \citep{Jiggens2012}. The paper is structured in 5 further chapters:
\begin{itemize}
\item Model underlying data and treatment (Section \ref{sec_mod_data});
\item Modelling methodology, statistical distributions and extrapolations (Section \ref{sec_model});
\item Model results including evolution with respect to previous work and rare SPE characteristics (Section \ref{sec_pro});
\item Discussion and comparisons with data and existing specification models of the SEP environment (Section \ref{sec_comp});
\item Concluding remarks (Section \ref{sec_conc}).
\end{itemize}

\section{Model data}\label{sec_mod_data}

\subsection{In-situ Data}\label{sec_model_data}

There are several problems apparent in SEP datasets, which require consideration when producing a dataset to be applied in an environment specification model and used as a reference for other studies. These issues include:
\begin{enumerate}
	\item The saturation or even paralysis of science-quality instrumentation during periods of high flux;
	\item Uncertainties in the response of monitor-quality data given broad energy bins;
	\item Data spikes and dead-time effects;
	\item The limited timespan of data being insufficient to characterise the variability of the SEP environment.
\end{enumerate}

The SAPPHIRE model is based on the second version of the SEPEM Reference DataSet (hereafter referred to as RDSv2) combining these data in a more homogeneous way than the previous version. In order to address these problems it was necessary to combine the different data sets available to take advantage of their best features. This includes cross-calibration of GOES/SEM(-2) radiation monitor data with IMP-8/GME data, which itself had to be corrected for a deterioration in performance and ultimate failure of the anti-coincidence detector between 1984 and 1990, as documented by \citet{Sandberg2014}. The crucial finding of this work was that the GOES proton channels (P2 - P7) when assigned the correct mean energy value exhibit an excellent linear correlation with data interpolated from the IMP-8/GME instrument (using only good data points). This energy calibration replaces previous calibration performed on the fluxes which may use corresponding calibration data from IMP-8/GME in the wrong energy range. Recent validation of derived integral proton fluxes from RDSv2 has been performed by \citet{Rodriguez2017}.

Data spikes, limited in the GOES data but much more numerous in the earlier SMS data, have largely been removed manually. RDSv2 includes extensively cleaned data from the SEM instruments on the SMS-1 and -2 and GOES-1, -2, -3 \& -5 spacecraft which represents an extension backwards in time of SEM data from 1986-01-01 to 1974-07-01 with respect to the previous version of the RDS (version 1.0). This has allowed the creation of a homogeneous and contiguous reference data set with a consistent processing chain spanning over 40 years (from 1974-2016) avoiding the need to use raw GME data, with its many gaps and saturated data points, as the basis for any time period.

The data is publicly available \citep{SEPEM_RDSv2} and the included readme file details the data merging and re-binning. The use of the different SEM(-2) data in the final RDS is shown in Table \ref{tab_semdata}. Data in the RDS are re-binned into the standard SEPEM energy channels shown in Table \ref{table_energies}.

\begin{table}[t]					
\centering	
\caption{\label{tab_semdata}Table of SEM(-2) data used in RDSv2. SEM-2 instruments flew on-board satellites from GOES-8 onward.}
\begin{tabular}{|c|c|c|} \hline 					
Spacecraft	&	Data Available	&	Usage in RDSv2	\\ \hline
SMS-01	&	1974-07-01 - 1975-10-31	&	1974-07-01 - 1975-01-31	 \\
SMS-02	&	1975-02-01 – 1978-03-31	&	1975-02-01 - 1977-03-31	 \\
GOES-01	&	1976-01-01 – 1978-05-31	&	1977-04-01 - 1977-07-31	 \\
GOES-02	&	1977-08-01 – 1983-05-31	&	1977-08-01 - 1983-05-19	 \\
GOES-03	&	1978-07-01 – 1979-12-31	&	(not used)	 \\
GOES-05	&	1984-01-01 – 1987-03-31	&	1983-05-20 - 1987-03-05	 \\
GOES-06	&	1983-05-01 – 1994-12-31	&	(not used*)	 \\
GOES-07	&	1987-03-01 – 1996-08-31	&	1987-03-06 - 1994-12-31	 \\
GOES-08	&	1995-01-01 – 2003-06-17	&	1995-01-01 - 2003-06-16	 \\
GOES-11	&	2000-07-01 – 2011-02-28	&	2003-06-17 - 2011-01-31	 \\
GOES-12	&	2003-01-01 – 2010-09-30	&	(not used)	 \\
GOES-13	&	2007-07-17 – 2015-12-31	&	2011-02-01– 2015-05-31	\\ \hline
\end{tabular}	
\\ \vspace{0.1cm}\small{*GOES-6/SEM data was used to fill 2 month of GOES-05/SEM data but no SPEs occurred.}								
\end{table}					

\begin{table}[t]							
\caption{\label{table_energies}Table of SEPEM reference energy channels and SPE flux thresholds (units for fluence are cm$^{-2}\cdot$sr$^{-1}\cdot$MeV$^{-1}$ and for peak flux are cm$^{-2}\cdot$sr$^{-1}\cdot$s$^{-1}\cdot$MeV$^{-1}$).}
\centering							
\begin{tabular}{|c|c|c|c|c|c|} \hline							
 & \multicolumn{3}{|c|}{Energy (MeV)} & \multicolumn{2}{|c|}{SPE model thresholds} \\							
Ch. 	&	 lower 	&	 upper 	&	 mean & Fluence  & Peak Flux \\ \hline
1	&	5.00	&	7.23	&	6.01 & 6.75E+5 & 1.00E+1	\\
2	&	7.23	&	10.46	&	8.70 & 3.80E+5 & 3.16E+0	\\
3	&	10.46	&	15.12	&	12.58	& 5.75E+4 & 1.00E+0 \\
4	&	15.12	&	21.87	&	18.18	& 1.30E+4 & 3.16E-1 \\
5	&	21.87	&	31.62	&	26.30	& 5.00E+3 & 1.00E-1 \\
6	&	31.62	&	45.73	&	38.03	& 1.20E+3 & 4.00E-2 \\
7	&	45.73	&	66.13	&	54.99	& 3.35E+2 & 2.50E-2 \\
8	&	66.13	&	95.64	&	79.53	& 2.00E+2 & 1.50E-2 \\
9	&	95.64	&	138.3	&	115.0	& 1.90E+2 & 6.00E-3 \\
10	&	138.3	&	200.0	&	166.3	& 6.00E+1 & 3.00E-3 \\
11	&	200.0	&	289.2	&	240.5	& 4.00E+1 & 2.00E-3 \\ \hline
\end{tabular}							
\end{table}							

\subsection{Solar Phase Definition}\label{sec_data_cyc}

The definition of a solar active period follows an assumed 7-year maximum in each (approximately) 11-year cycle nominally distributed 2.5 years before the maximum sunspot number and 4.5 year afterwards. This follows the definition used by \cite{Feynman1990a} who performed an superposed epoch analysis covering 3 solar cycles and found that solar proton fluences over 7-year periods offset from the timing of the sunspot maximum exceed those during the remaining 4 years on average by over an order of magnitude but that within this period the years of highest fluence varied. The definition of the cycle 23 solar maximum period has been modified from previous work to include the January 2005 SPE; the inclusion or exclusion of this SPE has a significant impact at high energies especially in the case of SPE peak fluxes. With this exception the model shows little sensitivity to small modifications of cycle definition. Table \ref{tab_cyc} shows the solar cycle active period start and end times.

\begin{table}[t]
	\centering
	\caption{Active periods for solar cycles 21-24.}
		\begin{tabular}{|c|c|c|c|c|} \hline
		Cycle & Start Date & End Date & SPEs & Mean Yearly Rate \\ \hline
		21 & 1977-05-26 & 1984-05-26 & 61 & 8.71 \\
		22 & 1987-05-26 & 1994-05-26 & 62 & 8.86 \\
		23 & 1998-03-01 & 2005-03-01 & 71 & 10.14 \\
		24 & 2009-01-01 & 2016-01-02 & 43 & 6.14 \\ \hline
		\end{tabular}
	\label{tab_cyc}
\end{table}

\subsection{Reference Event List}\label{sec_data_rel}

The SEPEM Reference Event List or REL (\url{http://sepem.eu/help/event_ref.html}) of SPEs follows the same definition as applied previously, requiring that the differential flux value in the 7.38-10.4 MeV channel is above 0.01 dpfu (dpfu = differential particle flux units = particles.cm$^{-2}\cdot$sr$^{-1}\cdot$s$^{-1}\cdot$(MeV/nuc)$^{-1}$), the minimum peak flux over the period is at least 0.5 dpfu, a dwell time of no more than 24 hours is permitted between consecutive enhancements (else they are treated as a continuation of the same SPE) and events must have a duration of at least 24 hours. The latest version of the REL includes 266 SPEs from between 1974 and 2016. The distribution of the 237 SPEs in the REL which occurred during active periods is shown in Table \ref{tab_cyc}, the remaining 29 SPEs occurred during solar minimum.

\subsection{Subtraction of background}\label{sec_data_bkg}

In-situ data shows a background level contributed by Galactic Cosmic Rays (GCRs) and instrument noise. The background level for each SPE in each channel has been taken to be equal to the mean of the flux in the three days before and after the event. This level has been subtracted from the flux profile for that channel in the SPE. The removal of the background results in a revision of the RDS referred to as RDSv2.1 \citep{SEPEM_RDSv2_1}. 

\begin{figure}
\centering
\includegraphics[width=0.65\columnwidth]{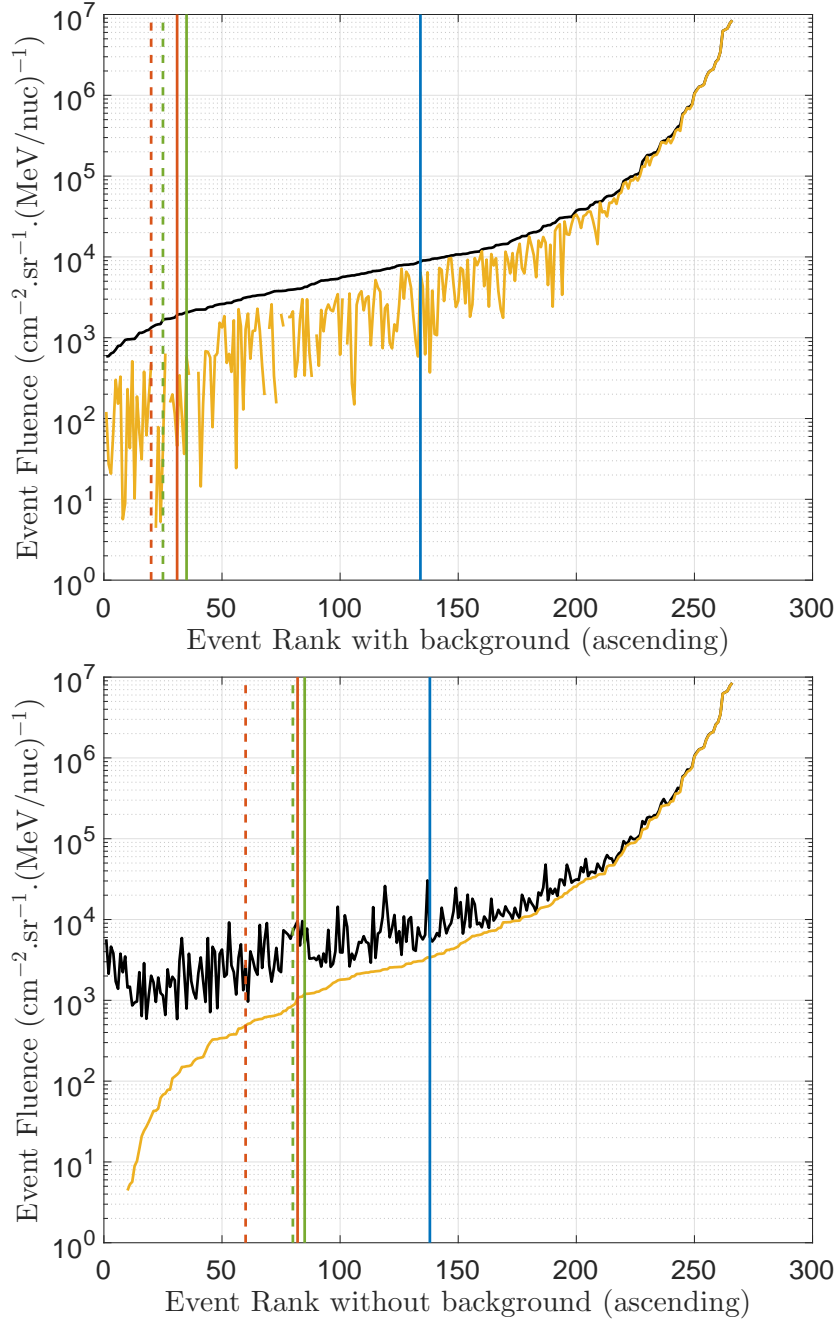}
\caption{Effect of background subtraction on SEP fluence for SEPEM Reference Energy Channel 6 (31.62 - 45.73 MeV). Black lines are the pre-background subtraction fluences while yellow lines are the fluences after subtraction of the background. The solid vertical lines correspond to the smallest event used in the optimised fitting distributions for the lognormal distribution (blue - Equation \ref{eqn_lognorm}), the truncated power law (red - Equation \ref{eqn_truncpl}) and the exponential cut-off power law (green - Equation \ref{eqn_copl}) as well as the effect of applying the minimum value as a free parameter in the fit for the two power laws (dashed lines - see text for details).}
\label{fig_bground}
\end{figure}

Figure \ref{fig_bground} shows the impact on SPE fluence of the background subtraction applied in SEPEM Reference Energy Channel 6 (31.62 - 45.73 MeV) for all SPEs in the SEPEM REL. The impact of the background subtraction reduces fluences by over an order of magnitude for the smallest SPEs in the list reflecting that no signal was detected in this channel for these events and that they must be excluded from the model. In the middle third of the 266 SPEs in the REL variations vary from a decrease of an order of magnitude to an almost negligible decrease due to the subtraction of background. This reflects that many longer duration SPEs have a high fraction of time spent at background levels while shorter duration SPEs can exhibit high energy fluxes for much of their duration. Finally, the largest 25 SPEs show almost no impact with the subtraction of background as the background flux is negligible in comparison to enhancements of up to 4 orders of magnitude such as seen on 20th October 1989. At higher energies the impact of the background subtraction extends to some of the largest SPEs while at lower energies only the smallest SPEs are impacted.

\section{Modelling Approach}\label{sec_model}

\subsection{Virtual Timelines Method}\label{sec_model_vtm}

The SAPPHIRE Modelling methodology is unchanged from the \textit{Virtual Timelines Method (VTM)} outlined by \citet{Jiggens2012}. The model generates SPEs are interspersed with waiting times (the time between events) sampled from a fitted distribution (the waiting time distribution) which is the Fourier transform of the event frequency distribution (see Section \ref{sec_model_wtd}).

\begin{figure}
\centering
\includegraphics[width=\columnwidth]{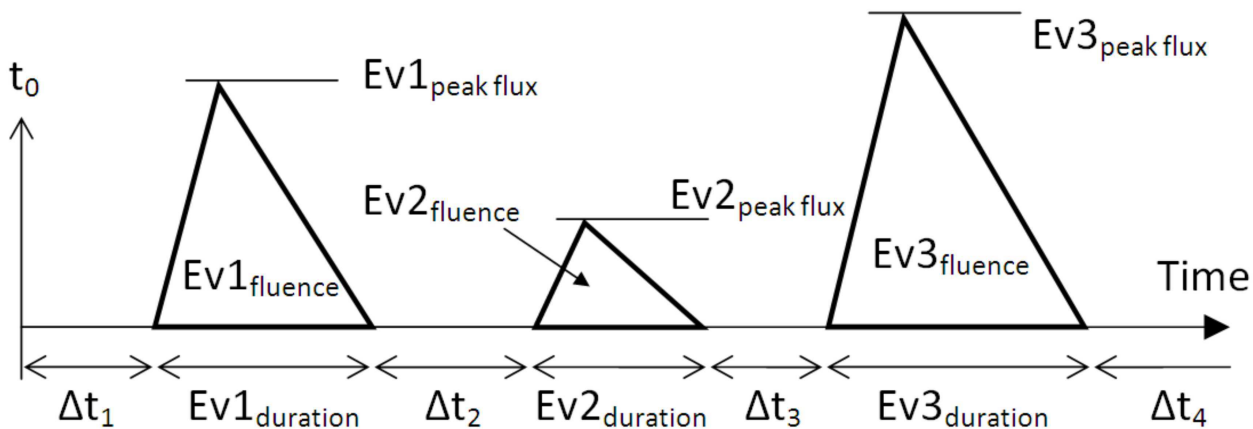}
\caption{Virtual Timelines Method (VTM) generating interspersed waiting times and SPEs with associated peak flux or fluence \citep{Jiggens2012}.}
\label{fig_timeline}
\end{figure}

Not all events in the REL show a signal in all channels. In order to include a single waiting time distribution for all channels and outputs, a portion of generated events are flagged as \textit{insignificant} in proportion to events in the REL which were below the flux threshold for that channel (see Section \ref{sec_model_opt} for details). These generated events are assigned a zero fluence and a duration sampled from a distribution based on the subset of events in the REL which were below the threshold. A separate distribution fit (described in Section \ref{sec_model_dist}) is made to SPE data at each of the 11 SEPEM reference energy channels which are sampled to generate an SPE flux (peak flux or fluence) for the remaining \textit{significant} SPEs which are generated. A numerical regression based on the flux level is used to determine a duration for each of these SPEs as described in \citet{Jiggens2012}. The main VTM code is then run for the prediction period (or mission length) requested. Figure \ref{fig_timeline} gives an illustration of a timeline generated with interspersed waiting times and events.

Once the run is complete (the mission duration reached) the highest SPE peak flux or the highest SPE fluence and cumulative fluence are recorded. The procedure is repeated for 100000 iterations and the outputs are ordered to derive the probability of exceeding (1 - confidence). This process is applied to all 11 energy channels separately. SAPPHIRE applies VTM to provide outputs of mission cumulative fluence, largest SPE fluence and peak flux for prediction periods up to 5 solar cycles for solar minimum and maximum separately. All outputs are given at 53 confidence intervals equal to a probability of exceeding ranging from 0.5 (50\% confidence) to 0.001 (99.9\% confidence).

Although more computationally intensive than previous Monte-Carlo methods, VTM allows for the inclusion of SPE durations and (although not yet implemented) the capability to alter the waiting time distribution of SPEs with progression through the phases of a solar cycle.

\subsection{Time distributions}\label{sec_model_wtd}

The waiting time distribution for solar maximum was based on a L\`{e}vy distribution as previously applied to solar flares \citep{Lepreti2001} and SPEs \citep{Jiggens2009}. The normalised L\`{e}vy complement cumulative distribution applied to SPE waiting times is given by:

\begin{equation}
	\centering
    P( \Delta t) = \frac{\exp(-|c\Delta t|^{\mu})}{\int^{\infty}_{t_{0}}\exp(-|c\Delta t|^{\mu})\textrm{d}\Delta t}
    \label{eqn_Levy}
\end{equation}

where $P( \Delta t)$ is the probability density for a waiting time $\Delta t$, $\mu$ is the characteristic exponent and $c$ is the scaling factor related to the mean frequency. The denominator normalises the fitted function to be equal to one over the range from $t_{0}$ to $\infty$. The value of $t_{0}$ was found to be 0.63 days which is close to the minimum dwell time of 1 day permitted in the event definition. The L\`{e}vy distribution is also applied to the SPE durations at each energy. During solar minimum there is an insufficient number of events (and therefore waiting times) to make a reasonable L\`{e}vy distribution fit. In this case the time-dependent Poisson distribution \citep{Wheatland2000,Wheatland2003} was applied to waiting times as it's single fitting parameter, $\varrho$, is related to the average event frequency which could be calculated simply but dividing the number of events occurring during solar minimum conditions by the the total number of days at solar minimum in the data set:

\begin{equation}
    P(\Delta t) = \frac{2 \varrho}{(1+\varrho\Delta t)^{3}}
    \label{eqn_TdP}
\end{equation}

\subsection{Flux Distributions}\label{sec_model_dist}

\begin{figure}
\centering
\includegraphics[width=0.9\columnwidth]{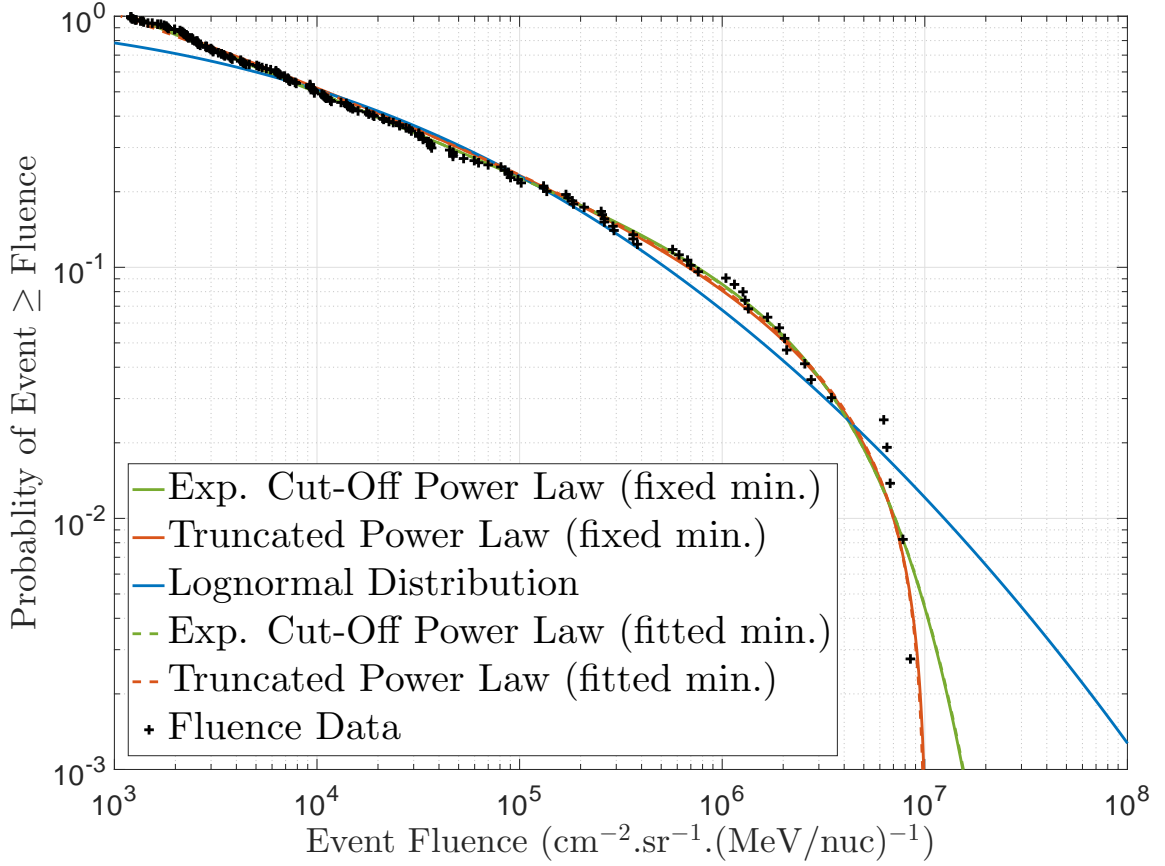}
\caption{Example flux distribution fits applied to SPE proton fluences in SEPEM Reference Energy Channel 6 (31.62 - 45.73 MeV).}
\label{fig_fludist}
\end{figure}

Three statistical distributions were investigated as part of this work to update the original VTM model described by \citet{Jiggens2012}. The first fitted function is the lognormal distribution (or more correctly the normal distribution applied to the base-10 logarithm of particle flux) used in the JPL model \citep{Feynman1993}:

\begin{equation}
\centering
    F\left(\phi\right)=1-\frac{1}{2}\left(1+\textrm{erf}\left[\frac{\log_{10}(\phi)-\mu}{\sigma\sqrt{2}}\right]\right)
    \label{eqn_lognorm}
\end{equation}

where $F\left(\phi\right)$ is the probability of a random event exceeding a fluence (peak flux), $\phi$, and $\mu$ and $\sigma$ are the mean and standard deviation of the $\log_{10}$ of the fluences (peak fluxes) respectively.
The second is the truncated power law applied in the ESP worst-case models \citep{Xapsos1998,Xapsos1999}:

\begin{equation}
\centering
    F\left(\phi\right)=1-\frac{\phi^{-b}_{\min}-\phi^{-b}}{\phi^{-b}_{\min}-\phi^{-b}_{\max}} =\frac{\phi^{-b}-\phi^{-b}_{\max}}{\phi^{-b}_{\min}-\phi^{-b}_{\max}}
    \label{eqn_truncpl}
\end{equation}

where $F\left(\phi\right)$ is the probability of a random event exceeding a fluence (peak flux), $\phi$, $b$ is the power law exponent, $\phi_{\min}$ is the minimum fluence (peak flux) and $\phi_{\max}$ is the maximum possible event fluence (peak flux) or `design limit'. The final distribution is the exponential cut-off power law introduced as part of the MSU model \citep{Nymmik2007}:

\begin{equation}
\centering
    F\left(\phi\right)=\frac{\phi^{-\gamma}}{\exp\frac{\phi}{\phi_{\lim}}} \frac{\exp\frac{\phi_{\min}}{\phi_{\lim}}}{\phi_{\min}^{-\gamma}} \approx\frac{\phi^{-\gamma}\phi_{\min}^{\gamma}}{\exp\frac{\phi}{\phi_{\lim}}}(\phi_{\lim}\gg\phi_{\min})
    \label{eqn_copl}
\end{equation}

where $F\left(\phi\right)$ is the probability of a random event exceeding a fluence (peak flux), $\phi$, $\gamma$ is the power law exponent, $\phi_{\min}$ is the minimum fluence (peak flux) and $\phi_{\lim}$ is the exponential cut-off parameter which determines the deviation from a power law at high fluences (peak fluxes). An example fit for these distributions applied to fluences of SPEs in the REL for the 6th energy channel is given in Figure \ref{fig_fludist}.

\subsection{Optimised Fitting}\label{sec_model_opt}

To derive the flux (fluence and/or peak flux) distributions for SAPPHIRE a study has been conducted to find the optimal minimum threshold (for each channel) for the inclusion of SPEs from the SEPEM REL. The results for the three distributions described above have been compared in this analysis.

The analysis begins with making a best fit to all of the 266 SPEs in the REL, recording the goodness-of-fit and removing the smallest SPE and making a new fit. The process is repeated until only 20 SPEs remain, with the aim of deducing the optimal number of SPEs to include within this energy range. Physically, the deviation from a distribution at low flux can result from too low signal-to-noise ratio due to the presence of instrument background (or GCRs) or the inability to detect smaller events in the presence of larger ones. Statistically, it is desirable to find a functional form and fitted parameters which best represent nature especially at high fluxes which will dominate model outputs at higher confidence levels usually applied in environment specifications.

The best fit is defined by the minimisation of the logarithm of the sum of squared residuals of the ordinate (probability). Although several fitting metrics were experimented with (including classical $\chi^{2}$ tests) this parameter gave the best balance of considering all events while giving more weight to the higher flux events and distribution fits which retained a larger proportion of the original 266 SPEs. The Goodness-of-Fit (GoF) was found by dividing the logarithm of the sum of squared residuals in the probability direction by the degrees-of-freedom (DoF; the number of SPEs minus the number of free parameters in the distribution). The number of free parameters for the lognormal distribution is two whereas the two power law distributions have three free parameters, however, the lognormal fit is made only to the top half of the SPEs above any given threshold (in keeping with descriptions given by \citet{Feynman1993}) drastically reducing the degrees-of-freedom with a maximum of: $266/2 - 2 = 131$. Where the goodness-of-fit parameter is lowest the best fit for each distribution is found and the indication of the threshold size for SPEs to be included in the model for that channel was given. This also allows a comparison between the distributions without biassing introduced by arbitrarily choosing the number of SPEs.

\begin{figure}
\centering
\includegraphics[width=0.85\columnwidth]{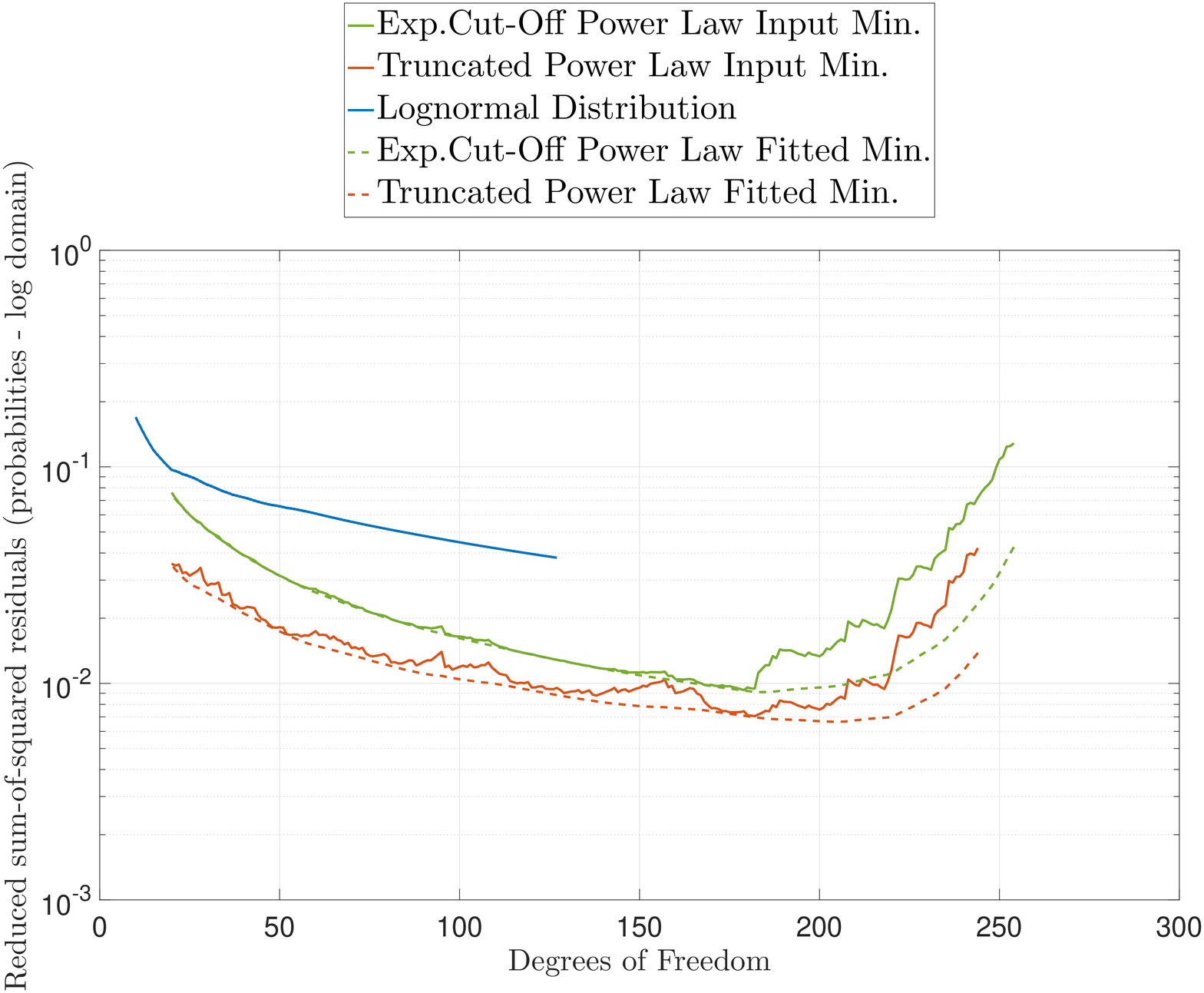}
\caption{Goodness-of-Fit variation as a function of degrees-of-freedom (the number of SPEs minus the number of free parameters in the distribution) for SPE proton fluences in SEPEM Reference Energy Channel 6.}
\label{fig_gof}
\end{figure}

For the two power law distributions (which include a minimum flux parameter) the impact of applying this as a free of fixed parameter (equal to the lowest flux above the threshold) was also studied. The benefit is that a better fit can be found, the negative aspect is that the distribution may predict a minimum event size above the size of events above the threshold. The results for the fluence distribution fits for Channel 6 of the RDSv2.1 are shown in Figure \ref{fig_gof}. The best results for this channel are achieved for the truncated power law ahead of the exponential cut-off power law while the lognormal distribution was a distant third. 

\begin{figure}
\centering
\includegraphics[width=\columnwidth]{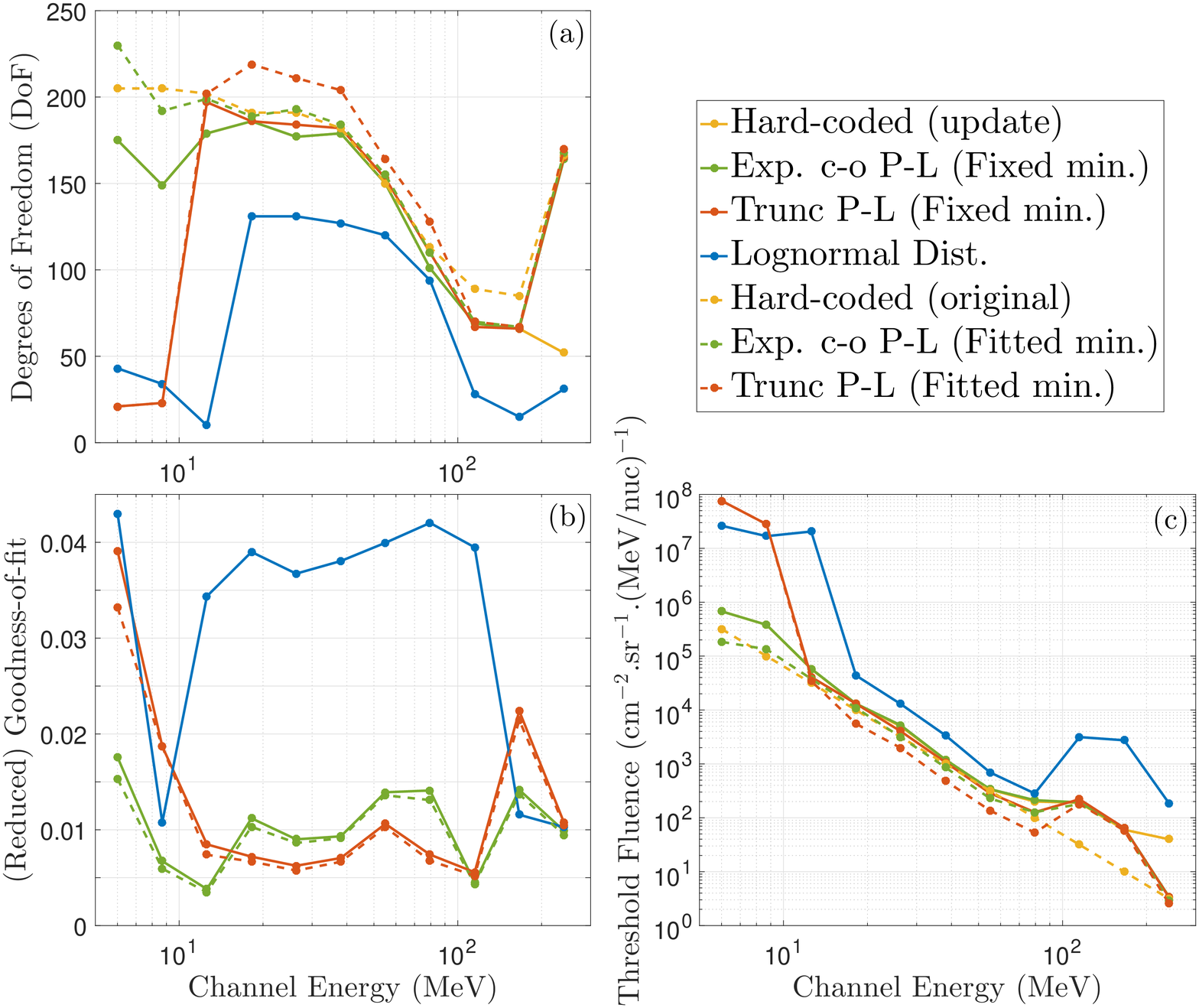}
\caption{Distribution fitting optimisation for SPE proton fluence as a function of energy for exponential cut-off power law (green), truncated power law (red) and lognormal distribution (blue). Top left panel: degrees of freedom for optimised solutions. Bottom left panel: (Reduced) goodness-of-fit parameters (see text for details). Bottom right panel: resulting thresholds to determine inclusion of SPEs from the REL.}
\label{fig_dof}
\end{figure}

Table \ref{tab:Gof_flu} and Figure \ref{fig_dof}(b) show the full set of results for SPE fluences. Although the truncated power law returns better GoF values in 5 energy channels it is outperformed by the exponential cut-off power law in the remaining 6 channels and the latter distribution gives a slightly lower mean value. The lognormal distribution is the best fitting in Channels 10 and 11 but is worse than the exponential cut-off power law by a factor of 3 overall. Peak flux GoF results (see Table \ref{tab:Gof_pkf}) show a similar result although now the truncated power law has the slightly lower mean. Unfortunately, this leaves the choice between the truncated power law and exponential cut-off power law to a great extent philosophical.

\begin{table}[t]
	\centering
	\caption{\label{tab:Gof_flu}Goodness-of-Fit (GoF) parameters for SPE proton fluence distribution fits: Exponential cut-off power law (ecopl - Equation \ref{eqn_copl}); Truncated power law (tpl - Equation \ref{eqn_truncpl}); lognormal distribution (lognorm - Equation \ref{eqn_lognorm}). The \textit{fix} and \textit{fit} denotes the treatment of the minimum fluence value (see text for details).}
		\begin{tabular}{|c|c|c|c|c|c|} \hline
Ch.	&	ecopl (fix)	&	tpl (fix)	&	lognorm	&	ecopl (fit)	&	tpl (fit)	\\ \hline
1	&	1.75E-02	&	3.90E-02	&	4.29E-02	&	1.53E-02	&	3.32E-02	\\
2	&	6.73E-03	&	1.87E-02	&	1.08E-02	&	5.94E-03	&	1.87E-02	\\
3	&	3.81E-03	&	8.47E-03	&	3.44E-02	&	3.47E-03	&	7.39E-03	\\
4	&	1.12E-02	&	7.19E-03	&	3.90E-02	&	1.03E-02	&	6.68E-03	\\
5	&	9.00E-03	&	6.24E-03	&	3.67E-02	&	8.62E-03	&	5.75E-03	\\
6	&	9.31E-03	&	7.05E-03	&	3.80E-02	&	9.11E-03	&	6.65E-03	\\
7	&	1.39E-02	&	1.06E-02	&	4.00E-02	&	1.36E-02	&	1.03E-02	\\
8	&	1.41E-02	&	7.40E-03	&	4.20E-02	&	1.31E-02	&	6.77E-03	\\
9	&	4.48E-03	&	5.57E-03	&	3.95E-02	&	4.31E-03	&	5.21E-03	\\
10	&	1.42E-02	&	2.24E-02	&	1.16E-02	&	1.37E-02	&	2.14E-02	\\
11	&	9.89E-03	&	1.08E-02	&	1.03E-02	&	9.46E-03	&	1.05E-02	\\ \hline
Mean	&	1.04E-02	&	1.30E-02	&	3.14E-02	&	9.72E-03	&	1.20E-02	\\ \hline
		\end{tabular}
\end{table}

\begin{table}[t]
	\centering
	\caption{\label{tab:Gof_pkf}Goodness-of-Fit (GoF) parameters for SPE proton peak flux distribution fits: Exponential cut-off power law (ecopl - Equation \ref{eqn_copl}); Truncated power law (tpl - Equation \ref{eqn_truncpl}); lognormal distribution (lognorm - Equation \ref{eqn_lognorm}). The \textit{fix} and \textit{fit} denotes the treatment of the minimum peak flux value (see text for details).}
		\begin{tabular}{|c|c|c|c|c|c|} \hline
Ch.	&	ecopl (fix)	&	tpl (fix)	&	lognorm	&	ecopl (fit)	&	tpl (fit)	\\ \hline
1	&	6.15E-03	&	3.46E-03	&	2.16E-02	&	5.83E-03	&	3.35E-03	\\
2	&	4.72E-03	&	2.56E-03	&	2.54E-02	&	4.29E-03	&	2.50E-03	\\
3	&	3.96E-03	&	2.13E-03	&	2.38E-02	&	3.62E-03	&	2.00E-03	\\
4	&	1.20E-02	&	7.19E-03	&	2.46E-02	&	1.12E-02	&	6.96E-03	\\
5	&	1.35E-02	&	8.62E-03	&	2.23E-02	&	1.26E-02	&	8.27E-03	\\
6	&	1.20E-02	&	8.06E-03	&	2.45E-02	&	1.12E-02	&	7.73E-03	\\
7	&	1.39E-02	&	1.18E-02	&	3.18E-02	&	1.33E-02	&	1.18E-02	\\
8	&	1.73E-02	&	1.63E-02	&	4.38E-02	&	1.68E-02	&	1.63E-02	\\
9	&	4.51E-03	&	6.26E-03	&	2.73E-02	&	4.37E-03	&	6.18E-03	\\
10	&	8.07E-03	&	1.17E-02	&	7.04E-03	&	8.04E-03	&	1.05E-02	\\
11	&	1.67E-02	&	1.66E-02	&	1.21E-02	&	1.58E-02	&	1.58E-02	\\ \hline
Mean	&	1.03E-02	&	8.61E-03	&	2.40E-02	&	9.73E-03	&	8.30E-03	\\ \hline
		\end{tabular}
\end{table}

When making a truncated power law fit to SPE fluences \citet{Xapsos1999} found a maximum event size, $\phi_{\max}$, significantly larger than the highest fluence SPE seen in their $>30$ MeV event list, they labelled this hypothetical maximum fluence SPE that the Sun could produce as seen from 1 AU as the `design limit.' However, the fits to the RDSv2 fluxes show that, at many energies, the truncated power law returns a maximum flux value close to the largest SPE in the REL. The exponential cut-off power law, on the other hand, has no such maximum size or `design limit.' The resulting distribution thereby allows for the possibility of larger events than have been seen to date (in the REL). It has been shown that the peak values measured by monitors at the near-Earth environment during October 1989 have been exceeded, for energies from 10-100 MeV, by those observed by STEREO-A during the SPE of July 2012 \citet{Jiggens2014}. For this reason, and because the summed mean values of peak flux and fluence are lower, the exponential cut-off power law was selected for the SAPPHIRE model. 

The analysis allowing the minimum parameter to vary (denoted \textit{fix} in Tables \ref{tab:Gof_flu} and \ref{tab:Gof_pkf}) for the two power laws (ecopl and tpl) shows a small reduction in the GoF but this was not implemented in the SAPPHIRE model as in some cases it removed the possibility of generating the smaller events which appear in the REL, however, it could be considered for any update. The derived SPE fluence thresholds applied in the model, shown in Figure \ref{fig_dof}(c), are taken from the best-fitting exponential cut-off power law fits (fixed minimum). The number of SPEs considered significant in each channel were based on these thresholds. Figure \ref{fig_dof}(b) shows that the thresholds selected resulted in $>50$ SPEs being considered even at the highest energies. Only the 11th energy channel threshold needed to be modified as the fitting procedure found a threshold with a higher DoF (see Figure \ref{fig_dof}(a)) than channels 7-10 and, on inspection, many of the flux profiles were at background levels during these SPEs. The final thresholds are shown by the solid yellow line (Hard-coded (update)) and in column 5 of Table \ref{table_energies}. More modifications had to be made for the peak flux thresholds which are shown in column 6 of Table \ref{table_energies}. With an update to the background subtraction routine it is hoped to fully automate this process in the future.

\subsection{1-in-x-year SPEs}\label{sec_pro_oix}

In addition to the dataset update (Section \ref{sec_model_data}) and the distribution fitting optimisation (Section \ref{sec_model_dist}), the major changes in the solar proton code are the inclusion of energy extrapolations to extend the energy range from 0.1 MeV to 1 GeV and the derivation of 1-in-x-year SPEs from the largest SPE fluence and SPE peak flux outputs.

For the largest SPE fluences and peak fluxes a method has been included to extrapolate the model outputs to give a spectrum for an SPE likely to occur an average of one time in a given number of years. The same SAPPHIRE model runs which produce a cumulative mission fluence for solar protons also provide the SPE fluence as a function of confidence, separate runs produce peak flux outputs. In the past this SPE has been labelled as the `worst-case' even though it is a function of prediction period and confidence. This concept is often difficult for non-experts to comprehend as the label `worst-case' implies something which will never be exceeded akin to the `design limit' in the early ESP models \citep{Xapsos1998,Xapsos1999}. However, what is meant is that there is a probability of $Y$\% that a given flux will not be exceeded by any SPE for a prediction period of $D$ years. For mission designers it stands to reason that the same confidence level as applied to the cumulative fluence would apply in this case. For other users of models it is more logical to think in terms of a `1-in-x-year SPE' - the fluence of SPE which will occur, on average, once in every $x$ years. 

In order to derive this spectrum it is assumed that very large events are distributed randomly in time following a Poisson distribution:

\begin{equation}
\textrm{Pr}⁡(N=k) = \frac{\lambda^{k} \exp(-\lambda)}{k!}  = \frac{1^{0} \exp(-1)}{0!} = 0.3679 = (N=0)
\end{equation}

where the mean value, $\lambda$, is set to 1 and the number of events for which the probability is calculated, $k$, is set to zero. This assumes that there is a model of the correct duration but as the goal is to derive spectra for very rare events the probability associated to an event which occurs on average once every $x$ years can be expressed as the cumulative probability for a shorter prediction period, $Pr_{D}$:

\begin{equation}
\textrm{Pr}⁡(N=0) = \textrm{Pr}⁡_{D}(N=0)^{\frac{x}{D}\times\frac{7}{11}}
\end{equation}

where $7/11$ is the average fraction of active years (only solar maximum model runs are used) and $D$ is the model prediction period (mission length) used. The SAPPHIRE model provides outputs as a function of the probability, $p$, of exceeding a given flux. The resulting quantity $Pr$ is the probability that a given flux is \textit{not} exceeded,($1 - p$), which is the same as the \textit{confidence} often quoted. This equation can then be rearranged to calculate $p$ for any given combination of $D$ and $x$:

\begin{equation}
\textrm{Pr}⁡_{D}(N=0) = \textrm{Pr}⁡(N=0)^{\frac{D}{x}\times\frac{11}{7}} = 0.3679^{\frac{D}{x}\times\frac{11}{7}} = 1-p
\end{equation}

The flux or fluence spectrum for $p$ is then used for an event that should occur, on average, once in $x$ years. Crucially, $D$ is linked to $p$ such that an infinite number of combinations of model duration and confidence are possible to obtain a desired result. The number of SAPPHIRE model runs executed for solar maximum conditions are only 21, but this was sufficient to verify the large SPE fluence outputs based on different model pairs of duration and probability. Where the required values of $p$, for given values of $D$, were within the the limits of the probabilities output stored by the model ($p>0.001$ and $p<0.5$) the outputs were found to be close to identical.

Outputs have been calculated for SPEs with an expected recurrence rate, $x$, of 1 in every 10, 20, 50, 100, 300, 1000 and 10000 years. Table \ref{tab_oix_select} shows the selected $p$ and $D$ pairs which have the smallest difference to the idealised values of $p(D)$ for 7 values of $x$. These outputs are equal to 4 decimal places in all but the 1-in-10-year SPE.

\begin{table}[t]
	\centering
	\caption{\label{tab_oix_select}Parameters for selection output runs as example 1-in-x-year SPEs.}
		\begin{tabular}{|c|c|c|c|} \hline
SPE	&	model	&	Prediction	&  Ideal\\
Freq. & Period ($D$) & Prob. ($p$) & $p(D)$ \\ \hline
10	&	2	&	0.2700 & 0.2697	\\
20	&	3	&	0.2100 & 0.2100	\\
50	&	3	&	0.0900 & 0.0900	\\
100	&	6	&	0.0900 & 0.0900	\\
300	&	18	&	0.0900 & 0.0900	\\
1000	&	26	&	0.0400 & 0.0400	\\
10000	&	32	&	0.0050 & 0.0050	\\ \hline
		\end{tabular}
\end{table}

\subsection{Spectral Extrapolations}\label{sec_pro_extrap}

The SAPPHIRE solar proton model is extended by making a Band Fit \citep{Band1993} to the output differential fluence spectrum from the model to extend the energies down to 0.1 MeV and up to 1 GeV. This spectral form has been increasingly applied to SPE fluence spectra so it seems appropriate to apply it also to the model outputs. \citet{Mewaldt2005} applied this formalism to spectra as a function of energy, however, \citet{Tylka2009} showed that for higher energies it is more correct to apply it to particle rigidity spectra which is how it is applied here. The Band function has 4 free parameters $C$, $R_{0}$, $\gamma_{a}$ and $\gamma_{b}$ and is given by:

\begin{align}\label{eq_band}\centering
\begin{split}
\textrm{d}J/\textrm{d}R = C\cdot R^{-\gamma_{a}} \textrm{exp}(-R/R_{0}) \hspace{3.0cm} & \Big\{ \textrm{for } R \leq (\gamma_{b}-\gamma_{a})R_{0} \Big\} \\
\textrm{d}J/\textrm{d}R = C\cdot R^{-\gamma_{b}}\Big(\left[ (\gamma_{b}-\gamma_{a})R_{0}\right]^{ (\gamma_{b}-\gamma_{a})}  \textrm{exp}(\gamma_{a}-\gamma_{b})\Big) \hspace{0.5cm} & \Big\{ \textrm{for } R \geq (\gamma_{b}-\gamma_{a})R_{0} \Big\}
\end{split}
\end{align}

where $R$ is the particle rigidity (momentum divided by charge) and $J$ is the SPE fluence or peak flux. Note that $(\gamma_{b}-\gamma_{a})R_{0}$ fixes the boundary between the two component functions of the Band fit and is constant, at 0.35 GV for all proton model outputs. Parameters $\gamma_{b}$ and $\gamma_{a}$ were tuned to ensure consistency between results for different confidence and prediction periods (no overlapping in the extrapolated region) while the scale parameter, $C$, was always left as a free parameter.

The Band fit was applied to 4 reference cases with tuned parameters to give spectral consistency between model outputs. For solar maximum these reference cases were:
\begin{enumerate}
\centering																
	\item prediction period: 3 yr; confidence: 75\%
	\item prediction period: 7 yr; confidence: 90\%
	\item prediction period: 20 yr; confidence: 95\%
	\item prediction period: 35 yr; confidence: 99\%
\end{enumerate}
Whereas for solar minimum these reference cases were:
\begin{enumerate}
\centering																
	\item prediction period: 10 yr; confidence: 75\%
	\item prediction period: 18 yr; confidence: 90\%
	\item prediction period: 32 yr; confidence: 95\%
	\item prediction period: 55 yr; confidence: 99\%
\end{enumerate}
These were chosen because they bound the region for most specification applications and give spectra spread evenly among the full set of output cases.

\section{Proton Model Results}\label{sec_pro}

\subsection{Mission Integrated Fluence}\label{sec_pro_mif}

\begin{figure}
\centering
\includegraphics[width=0.9\columnwidth]{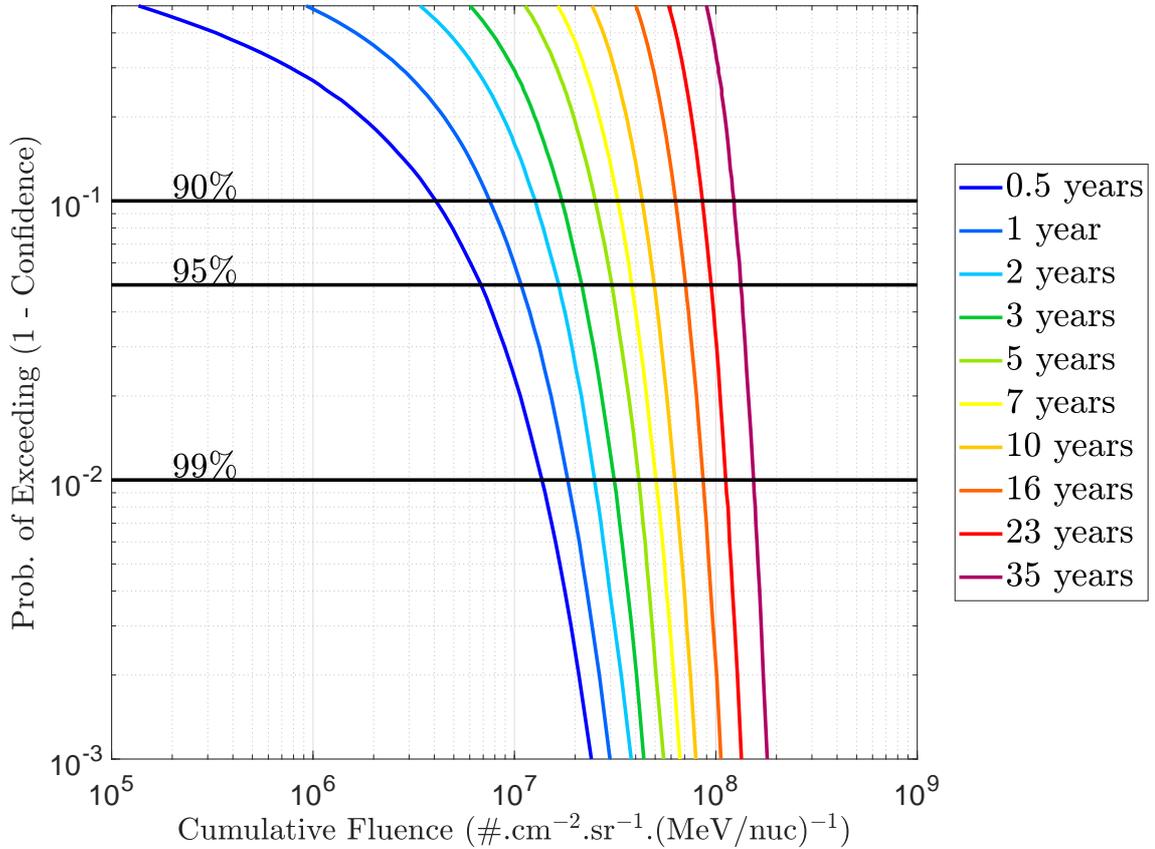}
\caption{SAPPHIRE proton cumulative fluence output for SEPEM reference Channel 6 (31.62 - 45.73 MeV).}
\label{fig_probflu}
\end{figure}

Model flux outputs are generated as a function of probability of exceeding the stated level. An example output of cumulative fluence plotted against probability of exceeding for Channel 6 at a set of mission durations during solar maximum is shown in Figure \ref{fig_probflu}. The outputs for each channel at a given confidence can be combined to produce a spectrum for specification purposes.

As described in Section \ref{sec_mod_data}, the processing chain for the SEPEM Reference Data Set (to produce the RDSv2) and Solar Particle Event (SPE) selection criteria has been updated significantly since publication of the original VTM by \citet{Jiggens2012}. Here the impact on model outputs of the following four updates are explored:

\begin{enumerate}
	\item The inclusion of earlier GOES and SMS data for the period from 1974-07-01 to 1986-01-01 in place of raw IMP-8/GME data creating a more homogeneous RDS with a fixed time resolution of 300 s (5 minutes) and an extension in time to cover all data up to 2015-12-31 extending the previous end date of 2009-06-30 with minor modifications to definition of solar active years in cycle 23;
	\item Update to the flux thresholds for inclusion of SPE at each channel as described in Section \ref{sec_model_opt} (original threshold values are given by the dashed yellow line in Figure \ref{fig_dof}(c);
	\item Updated data and processing given by corrections in IMP-8/GME data and cross-calibration of GOES/SEM/EPS based on uncertainty in the channel energies, as opposed to fluxes, resulting from the work of \citet{Sandberg2014};
	\item The subtraction of background fluxes described in Section \ref{sec_model_data}.
\end{enumerate}

\begin{figure}
\centering
\includegraphics[width=0.8\columnwidth]{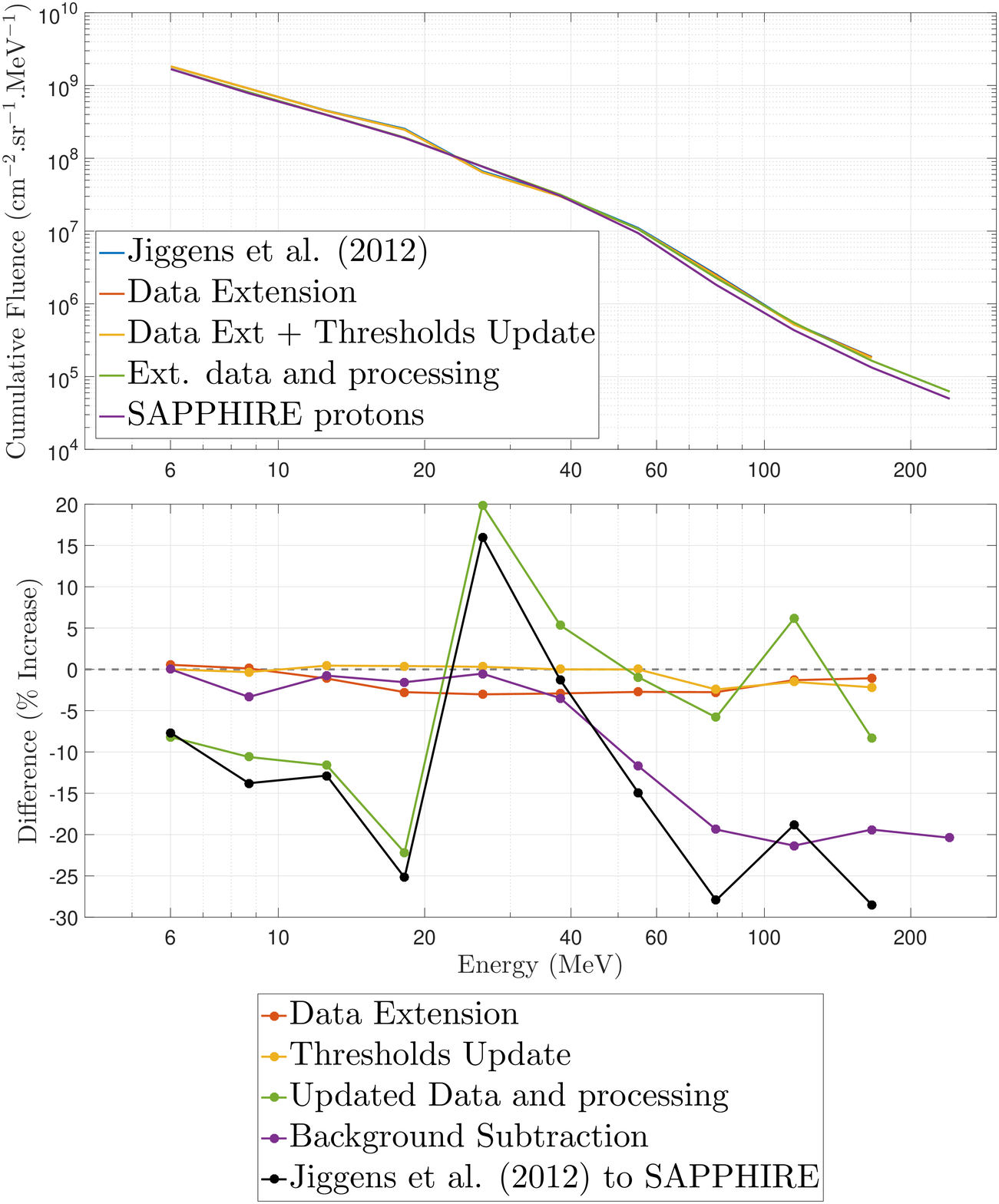}
\caption{Top: Model outputs for cumulative fluence at 95\% confidence and a prediction period of 5 years of solar maximum as a function of energy as each update (see text). Bottom: Model output evolution (percentage change) for cumulative fluence at 95\% confidence and a prediction period of 5 years of solar maximum as a function of energy.}
\label{fig_evo}
\end{figure}

The impact of these changes on the model results has been studied in detail for the cumulative fluence, largest SPE fluence and peak flux models. Figure \ref{fig_evo} (top) shows the outputs from the original VTM and after each of the updates listed above for the cumulative fluence model with a 95\% confidence level and a prediction period of 5 years of solar maximum as a function of energy. 

Each line in Figure \ref{fig_evo} (bottom) represents the percentage impact of the changes listed sequentially above plus the total change from the previously published model \citep{Jiggens2012} to the SAPPHIRE solar proton model. The range of variability is representative of the trends seen for all proton model outputs, confidence levels and prediction periods with the exception of the peak flux outputs for the top 3 energy channels where the inclusion of the January 2005 SPE has significantly impacted the high confidence results. The biggest differences result from the updated data and processing \citep{Sandberg2014} and, at higher energies, the subtraction of background. There is also a moderate reduction in output fluxes (2 - 3 \%) resulting from the inclusion of the rather quiet solar cycle 24 which also had few large SPEs. However, these changes in model output are modest in the context of such models and show remarkable model robustness given the modifications to data and fitting parameters. Note that the SPE of September 2017 is not included but an analysis showed that it was smaller than the 10 largest SPEs in the REL across all energies so the impact would not be dramatic. 

\begin{figure}
\centering
\includegraphics[width=\columnwidth]{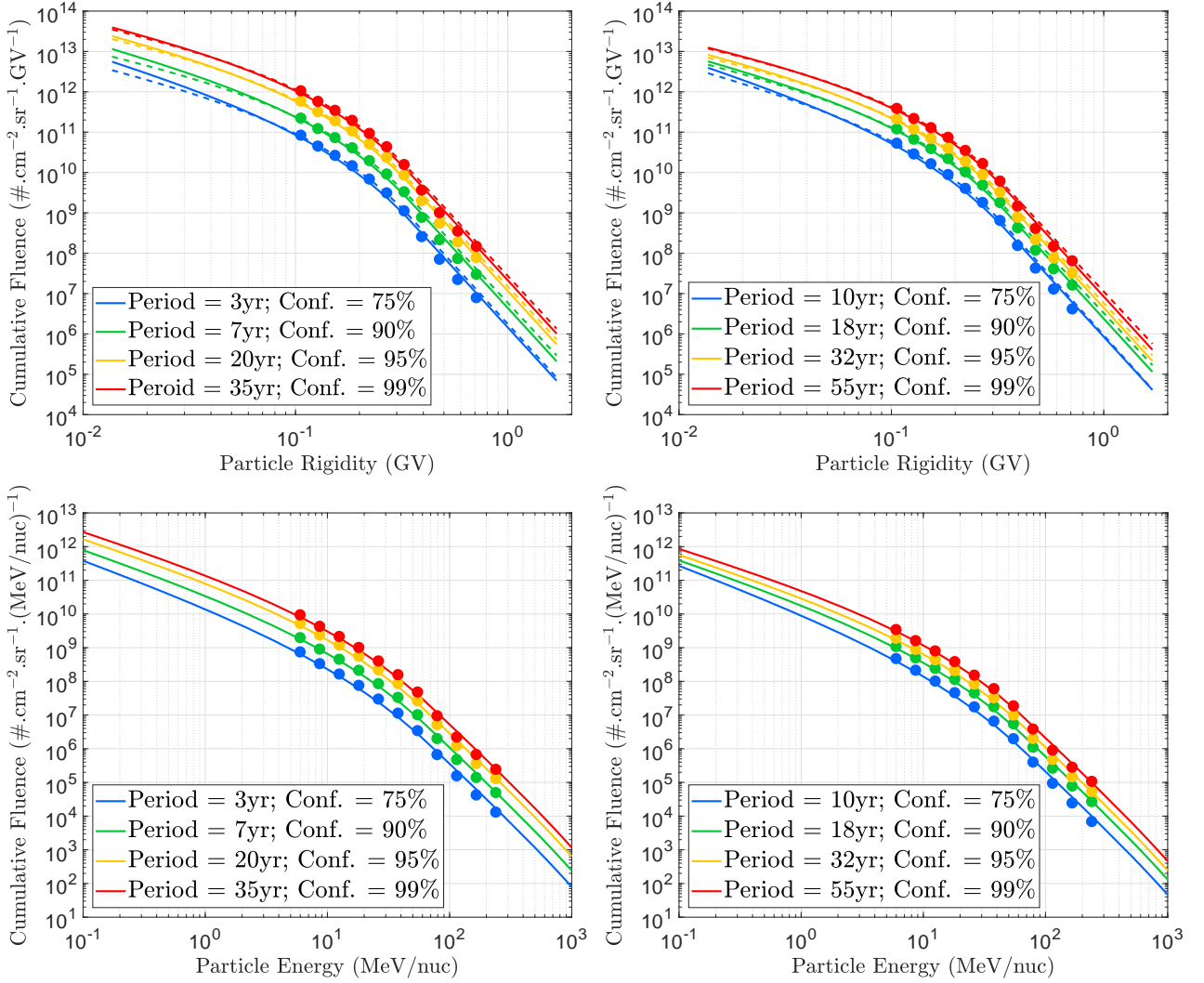}
\caption{Top: Band fits (dashed: original fits; solid: final fits - see text for details) to particle rigidity for proton cumulative fluence results at solar maximum (left) and solar minimum (right) for 4 reference cases. Bottom: Outputs expressed as a function of particle energy for solar maximum (left) and solar minimum (right).}
\label{fig_extrapR}
\end{figure}

\begin{figure}
\centering
\includegraphics[width=0.9\columnwidth]{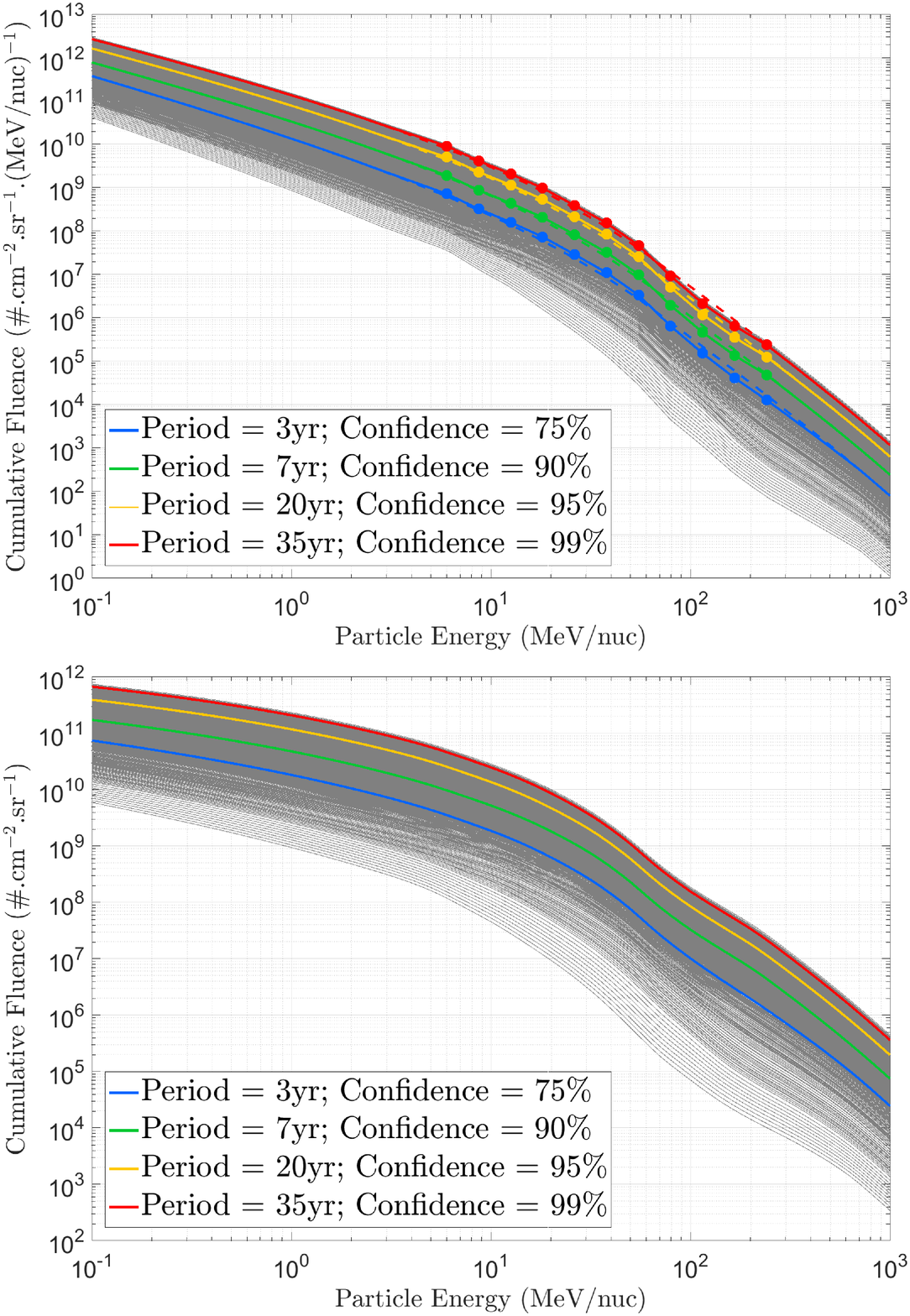}
\caption{Complete (extrapolated) spectra for solar proton cumulative fluence outputs at solar maximum differential in energy (top) and derived integral in energy (bottom). The thin grey lines show the complete set of model outputs calculated for 1113 pairs of prediction period and confidence.}
\label{fig_extrapE}
\end{figure}

Figure \ref{fig_extrapR} shows the original (best) fits made by the minimising the sum-of-square residuals in the log-domain (dashed lines) and the modified Band fits (solid lines) for cumulative fluence results at solar maximum and minimum. The final Band function fit parameters are shown in Table \ref{tab_band_pr} with numbering (1-4) corresponding to the reference cases given in Section \ref{sec_pro_extrap}. From the 4 fits made to the reference cases all other extrapolations were derived by interpolation/extrapolation linearly in log-space using Channel 1 and 11 outputs as boundary conditions. These results were then converted from spectral functions of particle rigidity back into functions of particle energy. The result for the solar proton cumulative fluence outputs at solar maximum is shown in Figure \ref{fig_extrapE} for differential in energy (top) and derived integral in energy (bottom). Note that within the 5 - 289 MeV range of the original model outputs the Band fits (dashed lines) are not used but instead a simple power law interpolation is applied (solid lines) in order to avoid modification of results where data is available. Results integral in particle energy use the power law index in rigidity at the highest energy in order to include the contribution of particles above 1 GeV.

\begin{table}[t]																
\caption{\label{tab_band_pr}Table of Band parameters used for SAPPHIRE proton models for mission cumulative fluence (cuflu), largest event time-integrated flux [fluence] (eiflu), largest event peak flux (epflu) for solar minimum and solar maximum conditions.}																
\centering																
\begin{tabular}{|c|c|c|c|c|c|c|c|} \hline 																
	&	\multicolumn{3}{|c|}{solar max.}					&		&	\multicolumn{3}{|c|}{solar min.}					 \\ 	
Parameter	&	cuflu	&	eiflu	&	epflu	&	Parameter	&	cuflu	&	eiflu	&	epflu	 \\ \hline	
$\gamma_{a,1}$	&	1.58	&	0.7	&	2.34	&	$\gamma_{a,1}$ &	1.66	&	0.8	&	2.5	 \\ 	
$\gamma_{a,2}$ 	&	1.41	&	0.44	&	2.02	&	$\gamma_{a,2}$ &	1.37	&	0.48	&	2.12	 \\ 	
$\gamma_{a,3}$  &	1.31	&	0.315	&	1.87	&	$\gamma_{a,3}$ &	1.26	&	0.34	&	1.97	 \\ 	
$\gamma_{a,4}$ &	1.26	&	0.19	&	1.74	&	$\gamma_{a,4}$ &	1.16	&	0.19	&	1.8	 \\ 	
$\gamma_{b}$	&	5.75	&	5.7	&	5.25	&	$\gamma_{b}$	&	5.75	&	5.7	&	5.25	 \\ 	
$R_{0,1}$ &	8.39E-02	&	7.00E-02	&	1.20E-01	&	$R_{0,1}$ &	8.56E-02	&	7.14E-02	&	1.27E-01	 \\ 	
$R_{0,2}$ &	8.07E-02	&	6.65E-02	&	1.08E-01	&	$R_{0,2}$ &	7.99E-02	&	6.71E-02	&	1.12E-01	 \\ 	
$R_{0,3}$ &	7.88E-02	&	6.50E-02	&	1.04E-01	&	$R_{0,3}$ &	7.80E-02	&	6.53E-02	&	1.07E-01	 \\ 	
$R_{0,4}$ &	7.80E-02	&	6.35E-02	&	9.97E-02	&	$R_{0,4}$ &	7.63E-02	&	6.35E-02	&	1.01E-01	 \\ 	
$C_{1}$ &	7.44E+09	&	2.57E+10	&	1.40E+04	&	$C_{1}$ &	3.72E+09	&	1.42E+10	&	6.39E+03	 \\ 	
$C_{2}$ &	3.20E+10	&	9.69E+10	&	6.66E+04	&	$C_{2}$ &	1.89E+10	&	6.51E+10	&	3.84E+04	 \\ 	
$C_{3}$ &	1.04E+11	&	1.94E+11	&	1.47E+05	&	$C_{3}$ &	4.41E+10	&	1.32E+11	&	8.38E+04	 \\ 	
$C_{4}$ &	2.11E+11	&	3.70E+11	&	2.91E+05	&	$C_{4}$ &	1.03E+11	&	2.89E+11	&	2.00E+05	\\ \hline	
\end{tabular}																
\end{table}																

\subsection{Rare SPE fluxes}

Band fits with particle rigidity were made for each of the 7 1-in-x-year SPEs in the same way to extrapolate the 1-in-x-year SPE model output spectra. The parameters found for SPE fluences and peak fluxes are displayed in Table \ref{tab_band_pr_oix}. The boundary $(\gamma_{b}-\gamma_{a})R_{0}$ and parameter $\gamma_{b}$ are retained from the values given in Table \ref{tab_band_pr} and again the scale parameter $C$ was always left as a free parameter.

\begin{table}[t]																			
\caption{\label{tab_band_pr_oix}Table of Band parameters used for SAPPHIRE proton 1-in-x-year SPE largest event time-integrated flux [fluence] (eiflu), largest event peak flux (epflu).}
\centering																			
\begin{tabular}{|c|c|c|c|c|c|c|c|c|c|} \hline 																			
	&	Param.	&	eiflu	&	epflu	&	Param.	&	eiflu	&	epflu	&	Param.	&	eiflu	&	epflu	 \\ \hline
1-in-10-year	&	$\gamma_{a}$	&	0.85	&	2.55	&	$R_{0}$	&	7.22E-02	&	1.30E-01	&	$C$	&	1.33E+10	&	6.03E+03	 \\ 
1-in-20-year	&	$\gamma_{a}$	&	0.71	&	2.35	&	$R_{0}$	&	7.01E-02	&	1.21E-01	&	$C$	&	2.81E+10	&	1.55E+04	 \\ 
1-in-50-year	&	$\gamma_{a}$	&	0.56	&	2.18	&	$R_{0}$	&	6.81E-02	&	1.14E-01	&	$C$	&	5.85E+10	&	3.58E+04	 \\ 
1-in-100-year	&	$\gamma_{a}$	&	0.48	&	2.09	&	$R_{0}$	&	6.71E-02	&	1.11E-01	&	$C$	&	8.73E+10	&	5.64E+04	 \\ 
1-in-300-year	&	$\gamma_{a}$	&	0.38	&	1.99	&	$R_{0}$	&	6.58E-02	&	1.07E-01	&	$C$	&	1.45E+11	&	9.60E+04	 \\ 
1-in-1000-year	&	$\gamma_{a}$	&	0.315	&	1.9	&	$R_{0}$	&	6.50E-02	&	1.05E-01	&	$C$	&	2.14E+11	&	1.53E+05	 \\ 
1-in-10000-year	&	$\gamma_{a}$	&	0.195	&	1.77	&	$R_{0}$	&	6.36E-02	&	1.01E-01	&	$C$	&	3.95E+11	&	2.96E+05	 \\ \hline
all	&	$\gamma_{b}$	&	5.7	&	5.25	&		&		&		&		&		&		 \\ \hline
\end{tabular}																			
\end{table}																			

\begin{figure}
\centering
\includegraphics[width=\columnwidth]{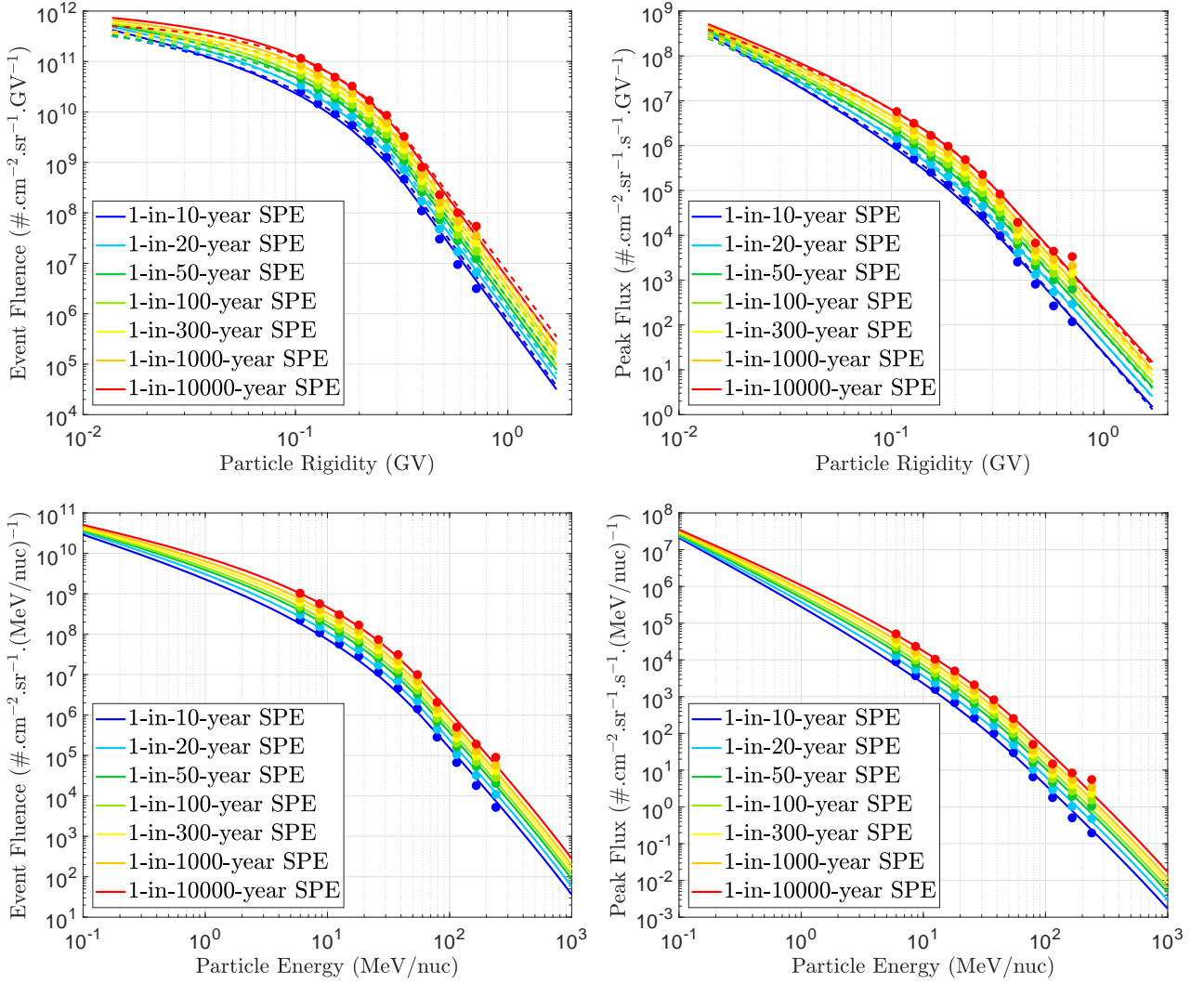}
\caption{Top: Band fits (dashed: original fits; solid: final fits - see text for details) to particle rigidity for proton 1-in-x-year event fluence (left) and peak flux (right) results. Bottom: Extrapolated spectra for SAPPHIRE 1-in-x-year event fluence (left) and peak flux (right) outputs as a function of particle energy.}
\label{fig_extrapOIX}
\end{figure}

Figure \ref{fig_extrapOIX} (top left) shows the Band fits for the proton 1-in-x-year event fluence calculations as a function of particle rigidity. The dashed lines represent the best fits through minimisation of the sum-of-squared residuals whereas the solid lines are the final fits applied to provide consistency between the resulting extrapolations. Below this is the same output transformed to a function of particle energy from 0.1 MeV to 1 GeV. Figure \ref{fig_extrapOIX} (top right) shows similar extrapolations for the proton peak flux calculations. Here there are excellent fits for all 1-in-x-year SPEs for all values below 66 MeV (channels 1-7), however, with increasing mean recurrence time there is increasing scatter in the fits. This may be due to the dominance of the SPE beginning on 15th January 2005 at the highest energies whereas other events dominate at lower energies creating a discontinuity in the spectrum. It may also be impacted by Channel 11 being extrapolated (rather than interpolated) from the GOES/SEM(-2) effective energies leading to higher uncertainty.

These 1-in-x-year SPEs are envelope spectra and unlikely to be produced by a single SPE which have very different spectral shapes from one another depending on characteristics of the CME and location on the solar disk. It is likely that much more data is needed to derive better fits for very rare SPEs. Unlike the other model outputs the 1-in-x-year SPE outputs within SAPPHIRE use the fitted spectra to smooth the impact of heavily extrapolated results based on models with high prediction periods and very low model probabilities (see Table \ref{tab_oix_select}).

\section{Discussion}\label{sec_comp}

\subsection{Integrated Fluence Model Comparisons}\label{sec_comp_cuflu}

\begin{figure}
\centering
\includegraphics[width=0.75\columnwidth]{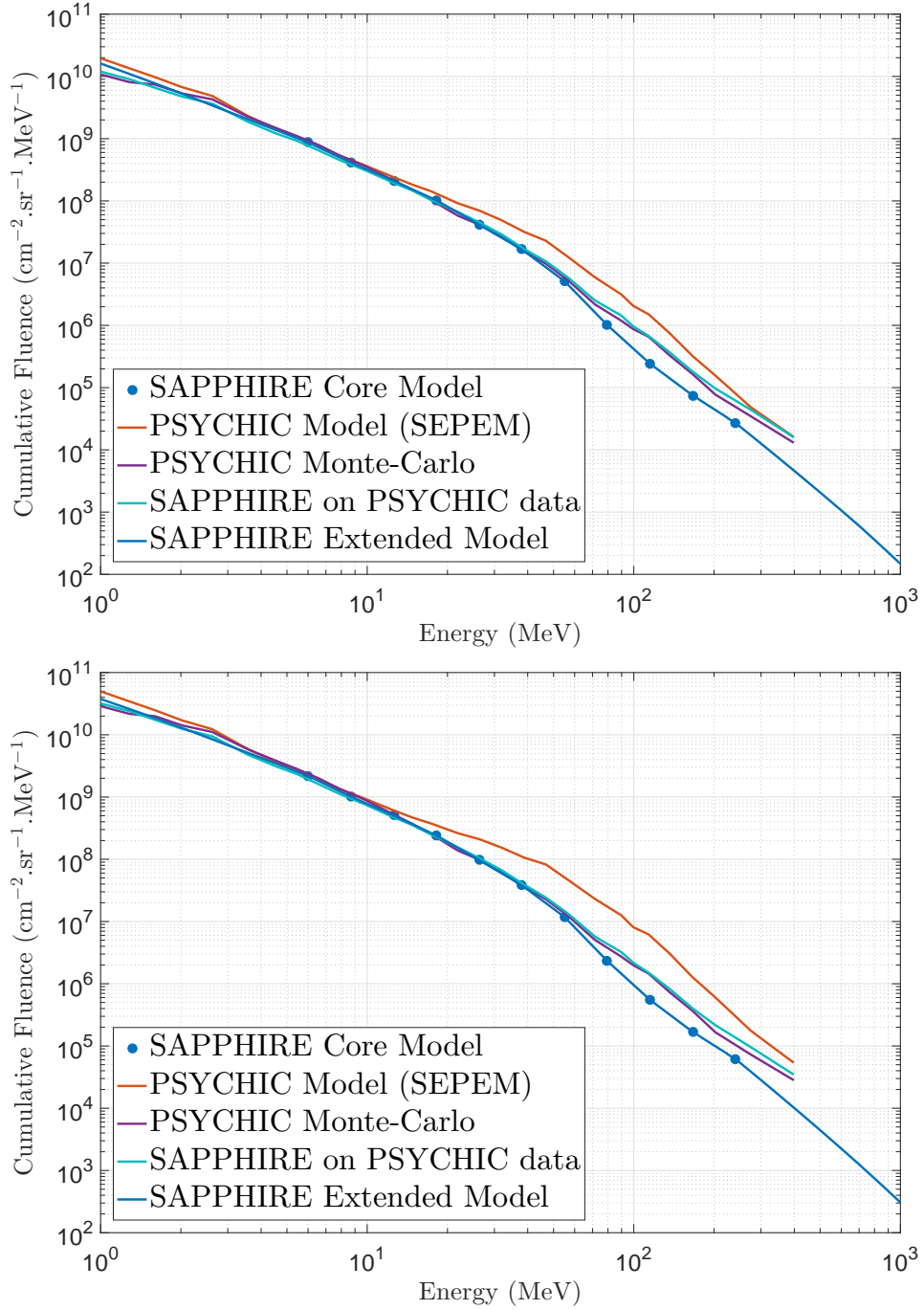}
\caption{Comparison of SAPPHIRE and PSYCHIC modelling approaches and data sets applied to a 2-year (top panel) and 7-year (bottom panel) solar proton cumulative fluence environment at a 95\% confidence level.}
\label{fig_comp1}
\end{figure}

The SAPPHIRE solar proton model cumulative fluence output at solar maximum has been compared to other models produced for specifying the SEP environment; namely the JPL \citep{Feynman1993} and PSYCHIC-ESP \citep{Xapsos2000} models. Some interesting conclusions can be gleaned from these comparisons. The first of these is that for short mission durations at nominal confidence levels (90\% or 95\%) the main difference in model outputs is driven by the treatment of the data. This is demonstrated in Figure \ref{fig_comp1} (top panel) where the SAPPHIRE model is compared to PSYCHIC-ESP applied only to the data in individual differential energy channels for a 2-year prediction period. This figure includes the output of the SAPPHIRE modelling approach applied to the PSYCHIC Integrated Data set (IDS) \citep{Xapsos2004} instead of the RDSv2.1 which shows a maximum deviation due to the modelling techniques of a factor of 2 at 50-100 MeV (compared to a factor of 3 differnce at 100 MeV for the two SAPPHIRE outputs changing only the dataset). Also plotted is a cumulative fluence model based on the truncated power law from the ESP worst-case fluence model \citep{Xapsos1999} combined with a Poisson distribution of event frequency following a JPL-type Monte-Carlo process (labelled PSYCHIC Monte-Carlo). This shows that for cumulative fluence models the choice between a truncated and exponential cut-off power law is not significant. The difference between the models due to data treatment is approximately a factor of 3 at 100 MeV but very low below 50 MeV. Figure \ref{fig_comp1} (bottom panel) shows the impact of extending the prediction period to 7 years. In this case the impact of the different modelling techniques has increased to a maximum of a factor of 4. This is because the ESP-PSYCHIC approach of fitting a lognormal distribution to the SEP yearly fluence increasingly diverges with respect to the SAPPHIRE VTM approach with increases in prediction period and confidence (see \citep{Jiggens2014}).

\begin{figure}
\centering
\includegraphics[width=0.75\columnwidth]{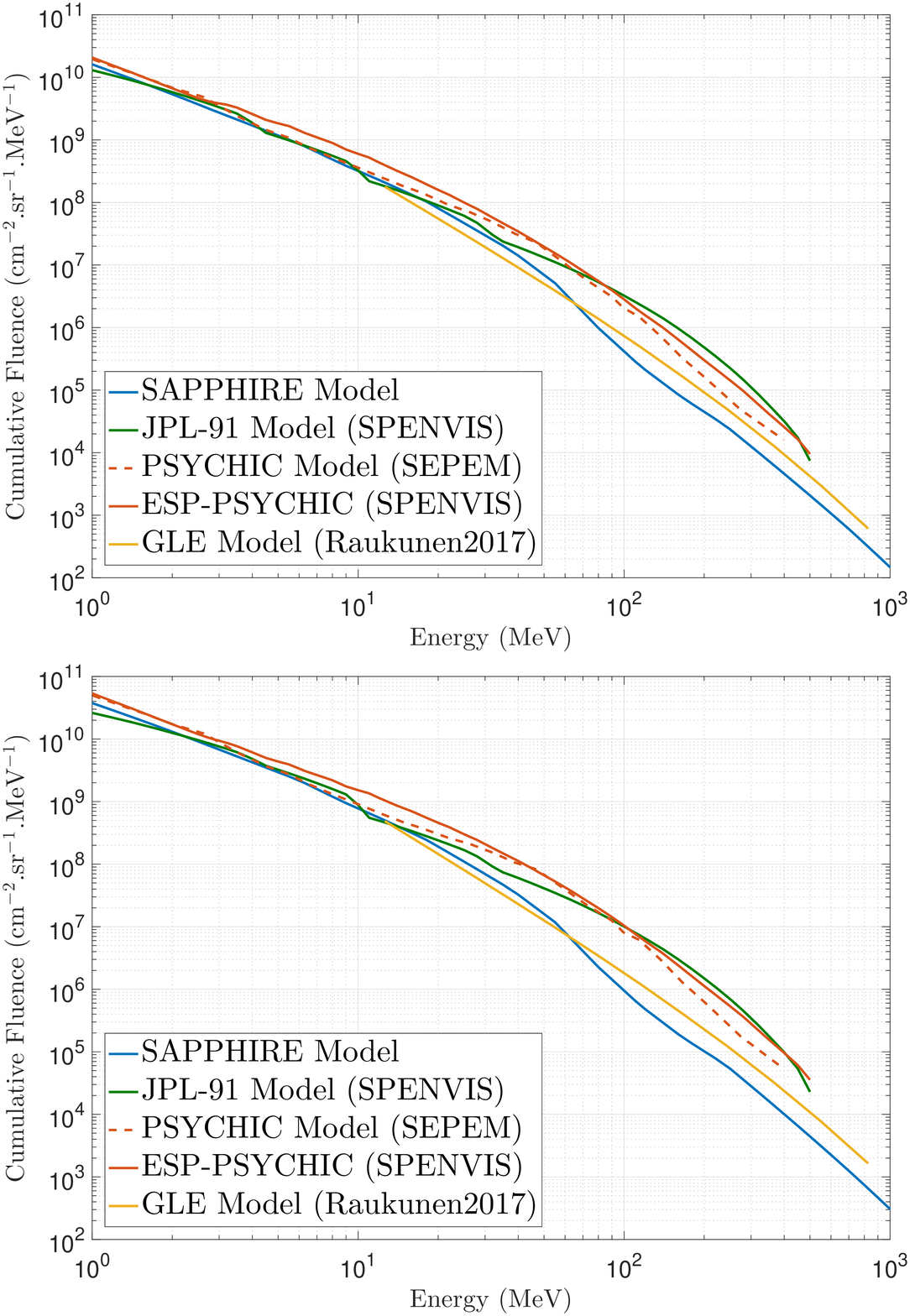}
\caption{Comparison of SAPPHIRE, PSYCHIC, JPL and GLE model outputs for a 2-year (top panel) and 7-year (bottom panel) solar proton cumulative fluence environment at a 95\% confidence level.}
\label{fig_comp2}
\end{figure}

Figure \ref{fig_comp2} shows the outputs from Figure \ref{fig_comp1} compared with the JPL model and PSYCHIC model as run on the SPENVIS system. These models have been run for integral fluence outputs and then differentiated within SPENVIS. In the case of PSYCHIC a fit is made to the integral spectra (see \citep{Xapsos2000}) prior to this differentiation. This shows that the way a model is implemented can have a significant impact on results. Despite good agreement between the models at 10 MeV the differences at 100 MeV are an order of magnitude. This is certainly of sufficient concern for calculation of effects in heavily shielded environments. A factor 3 difference between SAPPHIRE and PSYCHIC is attributable to the data differences at 100 MeV leaving a factor of 3 - 4 due to the modelling techniques. Based on previous work \citep{Jiggens2014}, the differences between SAPPHIRE and JPL are more attributable to the differences in modelling techniques than the model data.

Figure \ref{fig_comp2} also displays a new model produced by \citet{Raukunen2017} based on the characteristics of Ground Level Enhancements (GLEs) derived from Band fits to fluence spectra by \citet{Tylka2009} from Neutron Monitor (NM) observations. This model randomly samples parameters from Equations \ref{eq_band} with $C$ and $\gamma_{b}$ assumed to be normally distributed and with $\gamma_{a}$ and $R_{0}$ sampled based on linear regressions with $C$. The GLE model has a disadvantage at lower energies where fewer events are observed by Neutron Monitors (NMs) but a significant advantage at higher energies where the data span the energy range of interest and the data set extends over 5 solar cycles. The SAPPHIRE cumulative proton model output shows far better agreement with the GLE model than the JPL or ESP-PSYCHIC models over the total energy range. Nevertheless there are significant differences whic are explored further in Section \ref{sec_comp_dat}.

\subsection{SPE Fluence and Peak Flux Model Comparisons}\label{sec_comp_wcf}

\begin{figure}
\centering
\includegraphics[width=0.75\columnwidth]{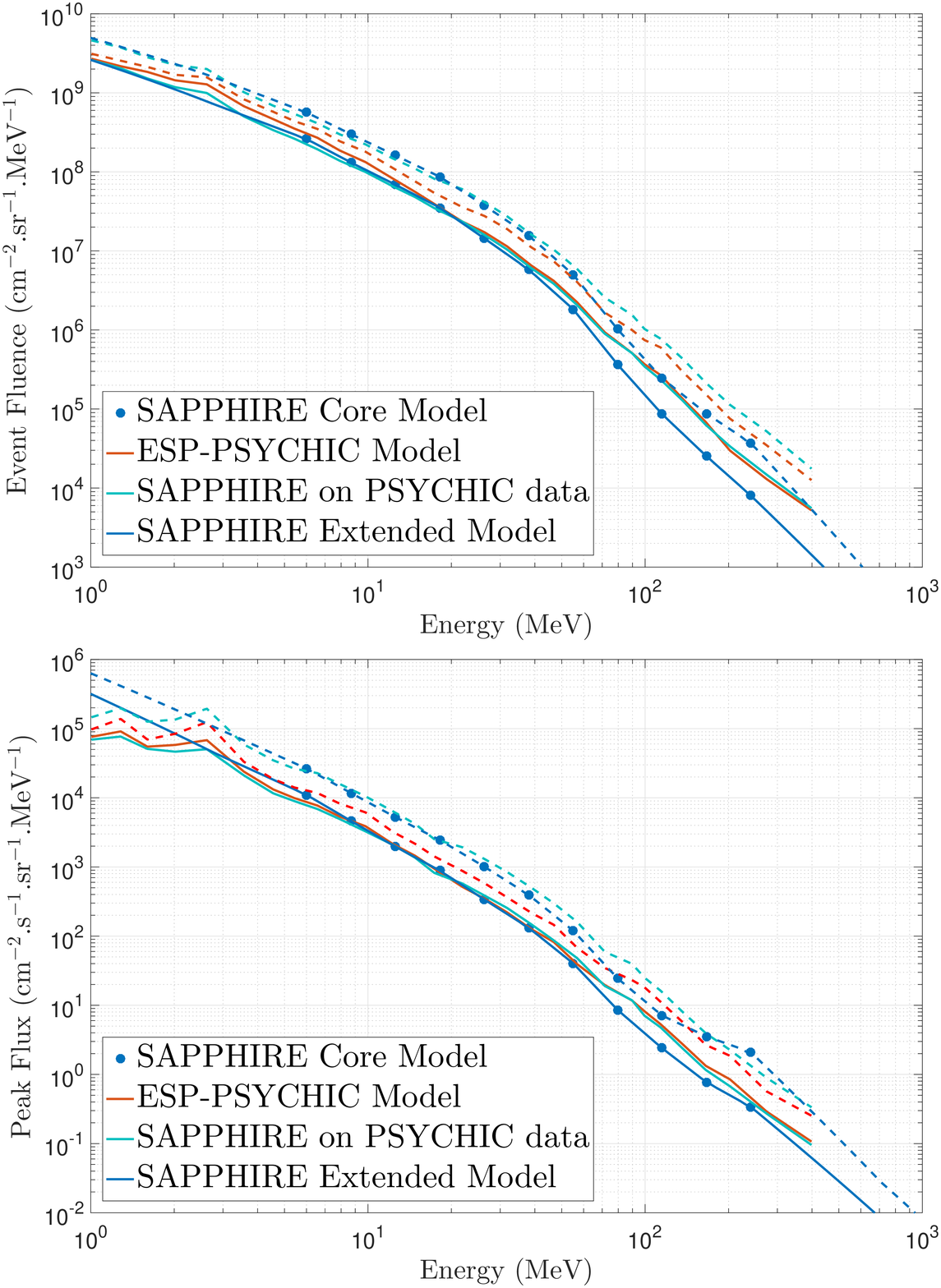}
\caption{Comparison of SAPPHIRE and ESP-PSYCHIC modelling approaches applied to worst-case SPE fluences (top) and peak fluxes (bottom) for 1 year at 90\% confidence (solid lines) and 7 years at 95\% confidence (dashed lines).}
\label{fig_compWCF}
\end{figure}

In a similar way to Figure \ref{fig_comp1}, Figure \ref{fig_compWCF} shows the differences in the SAPPHIRE model output compared with ESP-PSYCHIC for SPE fluence values (top) and peak flux values (bottom) differential in particle energy for two cases; 1-year prediction period at 90\% confidence (solid lines) and 7-year prediction period at 95\% confidence (dashed lines). Once more the SAPPHIRE approach has been additionally applied to the PSYCHIC data set in order to separate differences due to data and differences due to modelling methods. For the 1-year 95\% case the agreement up to 60 MeV is excellent and differences above this value are due to the different way the data have been processed which show more significant differences for SPE fluence than for peak flux. For higher prediction period and confidence the SAPPHIRE method returns higher values due to the differences in the power laws applied (see Figure \ref{fig_fludist}). For peak flux SAPPHIRE returns a higher output than the same model applied to the PSYCHIC data due to the presence of the January 2005 SPE in the SEPEM RDS which is the largest SPE in terms of peak fluxes. Although historically larger GLEs have been observed this event is the largest in terms of peak flux for three highest channels of the SEPEM RDS ($>95$ MeV) despite being only the 25th largest at 8.7 MeV (channel 2) and 10th largest at 26.3 MeV (channel 5). In the 11th SEPEM reference energy channel it has a peak flux a factor $\sim3$ larger than the next largest SPE.

\subsection{Data Comparisons}\label{sec_comp_dat}

\begin{figure}
\centering
\includegraphics[width=0.75\columnwidth]{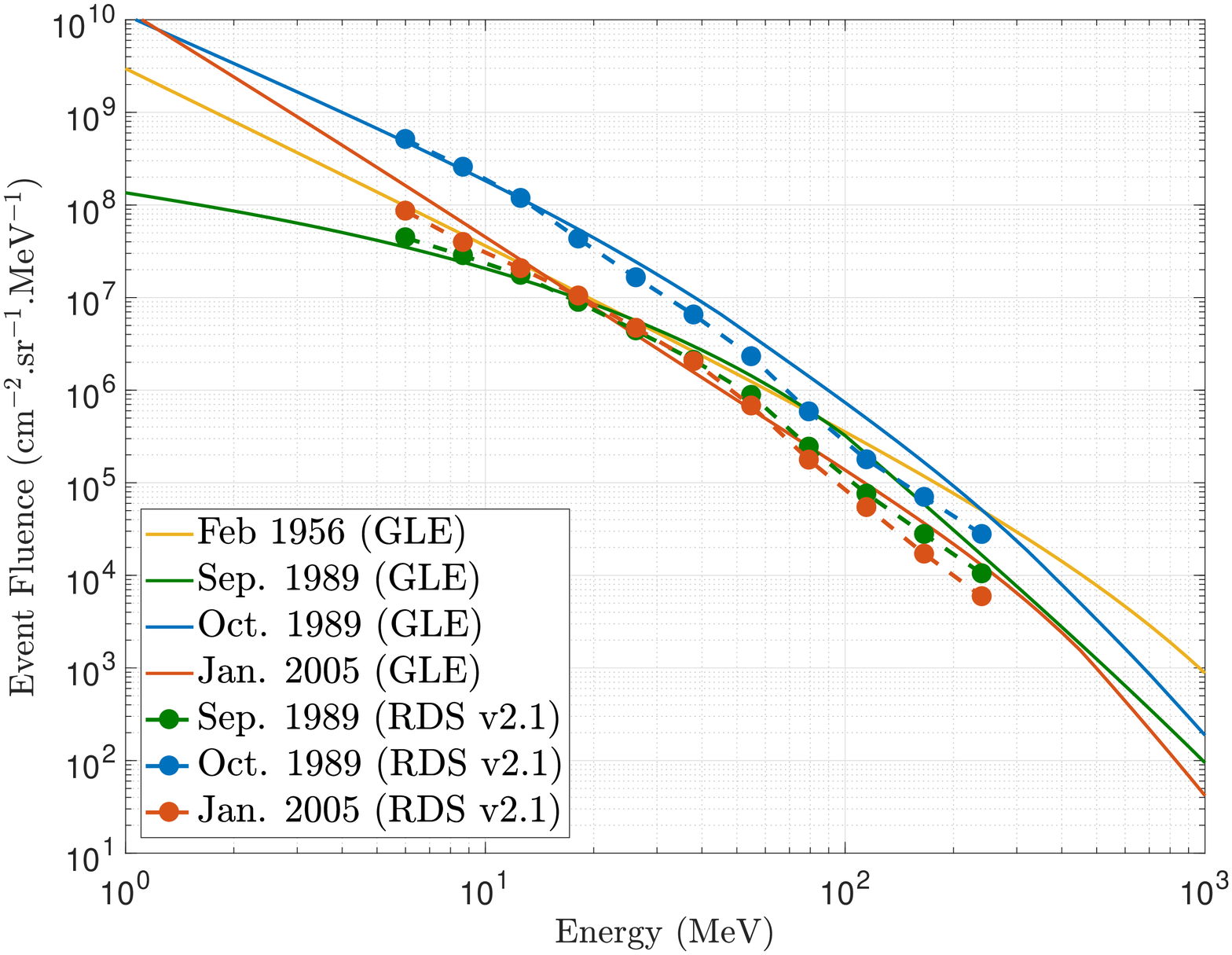}
\caption{Comparison of GLE-derived SPE fluence spectra with spectra from SEPEM RDSv2.1.}
\label{fig_NMcomp}
\end{figure}

The differences between the GLE fluence model of \citet{Raukunen2017} and SAPPHIRE below 65 MeV can be attributed to the contribution of smaller SPEs to the total fluence which are included in SAPPHIRE but not in the model based only on GLEs. For energies greater than 300 MeV the differences can be traced to the dominance of large GLEs observed in solar cycle 19 (notably in February 1956). Based on the GLE data the fluence of the February 1956 (GLE 5; GLE Episode 1) SPE exceeds that of October 1989 (GLE 43 \& 43, ESP 44 \& 45; GLE Episode 32) by a factor of 5 (see Figure \ref{fig_NMcomp}). The October 1989 SPE is the largest event in the SEPEM REL for fluences $>80$ MeV based on data from the SEPEM RDSv2.1, so the inclusion of earlier SPEs in the GLE fluence model could explain the model differences at the highest energies. However, this does not adequately explain the differences between the models at energies from 65 - 300 MeV where the February 1956 GLE is not as dominant. In this region the differences in the derived spectra for the September 1989, October 1989 and January 2005 SPEs show that the GLE fits return higher values than the data from the SEPEM RDSv2.1. This stems from the space-based data used by \citet{Tylka2009} to complete the spectral fit from the NM data which had been validated using high energy GOES/HEPAD, IMP8 and SAMPEX data. These data included data from GOES/SEM/EPS/MEPAD which is the same data used for SAPPHIRE. However, these data have been corrected in the RDS v2.1 as explained in Section \ref{sec_model_data} giving lower effective mean energy values for the higher energy channels. This would explain the differences in Figure \ref{fig_NMcomp}. It is important that these differences are resolved to verify cumulative fluence proton results for energies $>65$ MeV.

\begin{figure}
\centering
\includegraphics[width=0.9\columnwidth]{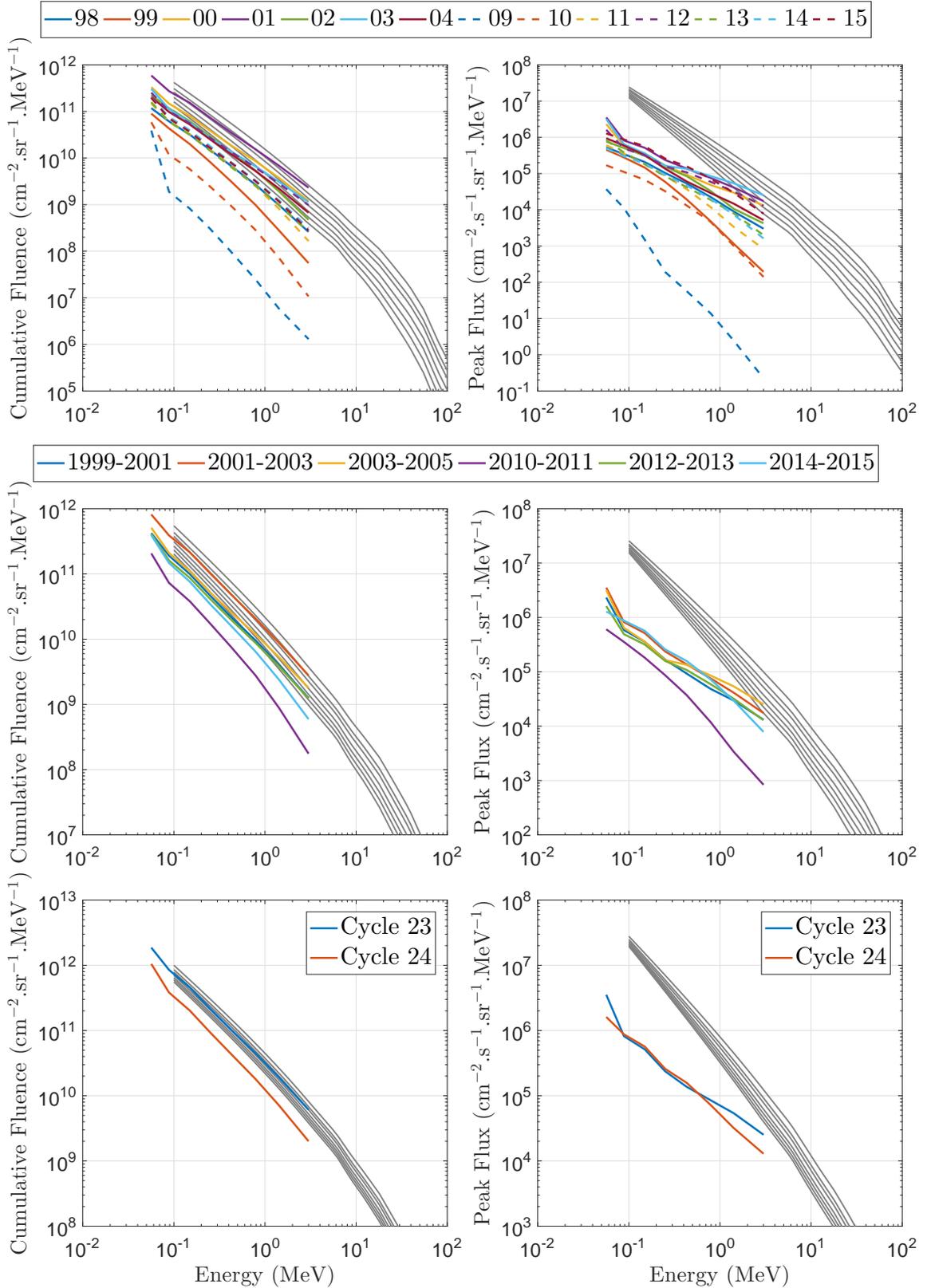}
\caption{Comparison of SAPPHIRE outputs (at confidence levels of 50\%, 60\%, 70\%, 80\%, 90\%, 95\% and 99\% - grey lines) for proton cumulative fluence (left) and peak flux (right) at solar maximum with data from ACE/EPAM for 1 year (top panel), 2 years (middle panel) and 7 years (bottom panel).}
\label{fig_compEPAM}
\end{figure}

Having investigated the comparison of the SAPPHIRE proton model outputs with a model based on GLEs it is also interesting to consider the appropriateness of the extrapolated low energy component of the model output. For this we take solar ion data from ACE/EPAM:\\
\url{http://www.srl.caltech.edu/ACE/ASC/level2/lvl2DATA_EPAM.html}\\
We have assumed that these data are dominated ($>95\%$) by solar protons and have not attempted to correct for caveats such as electron contamination. The solar maximum years corresponding to the definition in SAPPHIRE have been extracted for the comparison providing 14 years at solar maximum which have been binned in sets of 2 consecutive years (3 per cycle) and 7 years (complete cycle). Figure \ref{fig_compEPAM} shows the comparisons of with the model outputs for 50\%, 60\%, 70\%, 80\%, 90\%, 95\% and 99\% confidence intervals (grey lines). For the cumulative fluence outputs the agreement appears to be very good, 8 of the years are below 50\% confidence (reflecting the weakness of cycle 24) and 6 above. One point for concern might be that at energies above 1 MeV the ACE/EPAM data for the year 2001 exceeds the 95\% confidence level reaching almost 99\% at 3 MeV. This was a very active year however, so it appears reasonable that it might approach these levels. For the peak flux there is reasonable agreement at 3 MeV although the EPAM data is low compared to model outputs. Of greater concern is the extrapolation down to 0.1 MeV where there is an order of magnitude difference. This implies that the extrapolations for peak flux may be too harsh. However, it should be noted that the peak values for the lowest energy channel in SAPPHIRE come from events in cycle 22 and that surprisingly the most severe year in the EPAM data at 1 MeV is 2015 which is the last year of a weak cycle (24) principally due to the large but soft SPE on 21st June 2015. Further investigation of the data is needed to ensure that there are no caveats before using these data to justify any modification of the spectral extrapolation for SAPPHIRE. For the time being it can be assumed that SAPPHIRE is conservative in terms of low energy peak fluxes.

\subsection{Comparisons with CREME96}\label{sec_comp_wc}

\begin{figure}
\centering
\includegraphics[width=0.9\columnwidth]{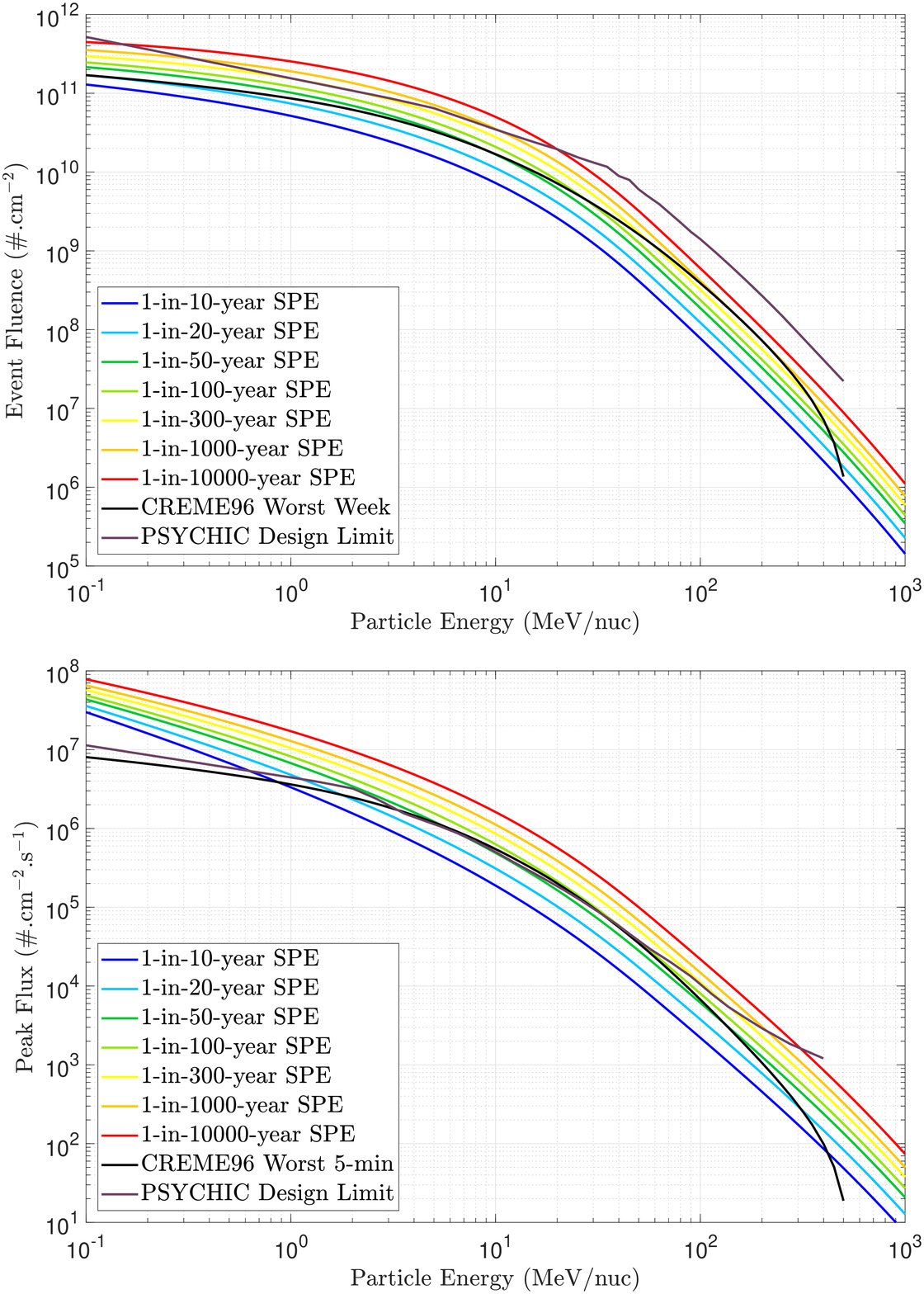}
\caption{Comparison of SAPPHIRE 1-in-x-year SPE fluences (top) and peak fluxes (bottom) with CREME96 worst week and the ESP-PSYCHIC `Design Limit'.}
\label{fig_compOIX}
\end{figure}

Figure \ref{fig_compOIX} (top) shows the derived SPE integral fluence values for the 1-in-x-year SPEs compared to an approximation of the ESP-PSYCHIC Worst Case (or `Design Limit') taken from a run on SPENVIS applying a probability of 99.999\% and the CREME96 Worst Week output. This shows that the current CREME96 output is approximately equal to a SAPPHIRE model 1-in-20 to 1-in-50-year output for energies below $10$ MeV, at $100$ MeV the CREME96 spectrum is between a 1-in-300-year SPE and 1-in-1000-year SPE and at $>500$ MeV the CREME96 spectrum drops below the 1-in-20-year SPE. It would be expected that the October 1989 SPE would be equivalent to a 1-in-40-year SPE given the duration of the RDS. The reason for the divergence from $20$ to $450$ MeV are due to the treatment of the GOES/SEM/EPS/MEPAD data in the derivation of the SEPEM RDS and the spectral form applied in CREME96. This gives rise to differences in the flux of approximately a factor of 2 at high energies a reduction of which would reduce the CREME96 spectrum to below the 1-in-100-year SPE level and close to the level anticipated by the SAPPHIRE 1-in-50-year SPE. At the lowest energies the ESP-PSYCHIC SPE fluence `Design Limit' is approximately equal to a 1-in-100-year SPE and at $>10$ MeV it is approximated by a 1-in-1000-year SPE. However, due to differences in data treatment it exceeds even the 1-in-10000-year SPE for (integral) energies greater than 20 MeV and at energies greater than 100 MeV this difference is approximately a factor of two.

The result for peak flux 1-in-x-year SPE from the SAPPHIRE model is compared to CREME96 and ESP-PSYCHIC in Figure \ref{fig_compOIX} (bottom). PSYCHIC peak flux `Design Limit' values have been calculated on the SEPEM system using the PSYCHIC IDS \citep{Xapsos2004} as this output is not available on SPENVIS. The CREME96 worst 5-minute output is approximately equivalent to a 1-in-100-year SPE at $>30$ MeV, the proton energy needed to penetrate nominal spacecraft shielding. CREME96 was based on the October 1989 SPE in order to avoid unrealistically severe environment models \citep{Tylka1997a} into the spacecraft design process and for these energies it appears to achieve that goal although once more the CREME96 result is enhanced with respect to expectations due to the data treatment. However, at high energies other SPEs in the SEPEM REL (notably the SPE of January 2005) have comparable or elevated fluxes in comparison to that of October 1989 so the CREME96 output is no longer conservative above $>100$ MeV. The ESP-PSYCHIC `Design Limit' for peak flux agrees closely with CREME96 for the lower part of the energy range and is approximately equal to a 1-in-100-year SAPPHIRE SPE between 5 and 70 MeV. The differences at low energies are due to different extrapolations used in each case. Divergence at higher energies results in the ESP-PSYCHIC `Design Limit' exceeding the 1-in-10000-year SPE at 400 MeV, however for use as a `worst-case' the agreement is good.

\subsection{Cumulative Fluence Model Progression}\label{sec_comp_prog}

\begin{figure}
\centering
\includegraphics[width=\columnwidth]{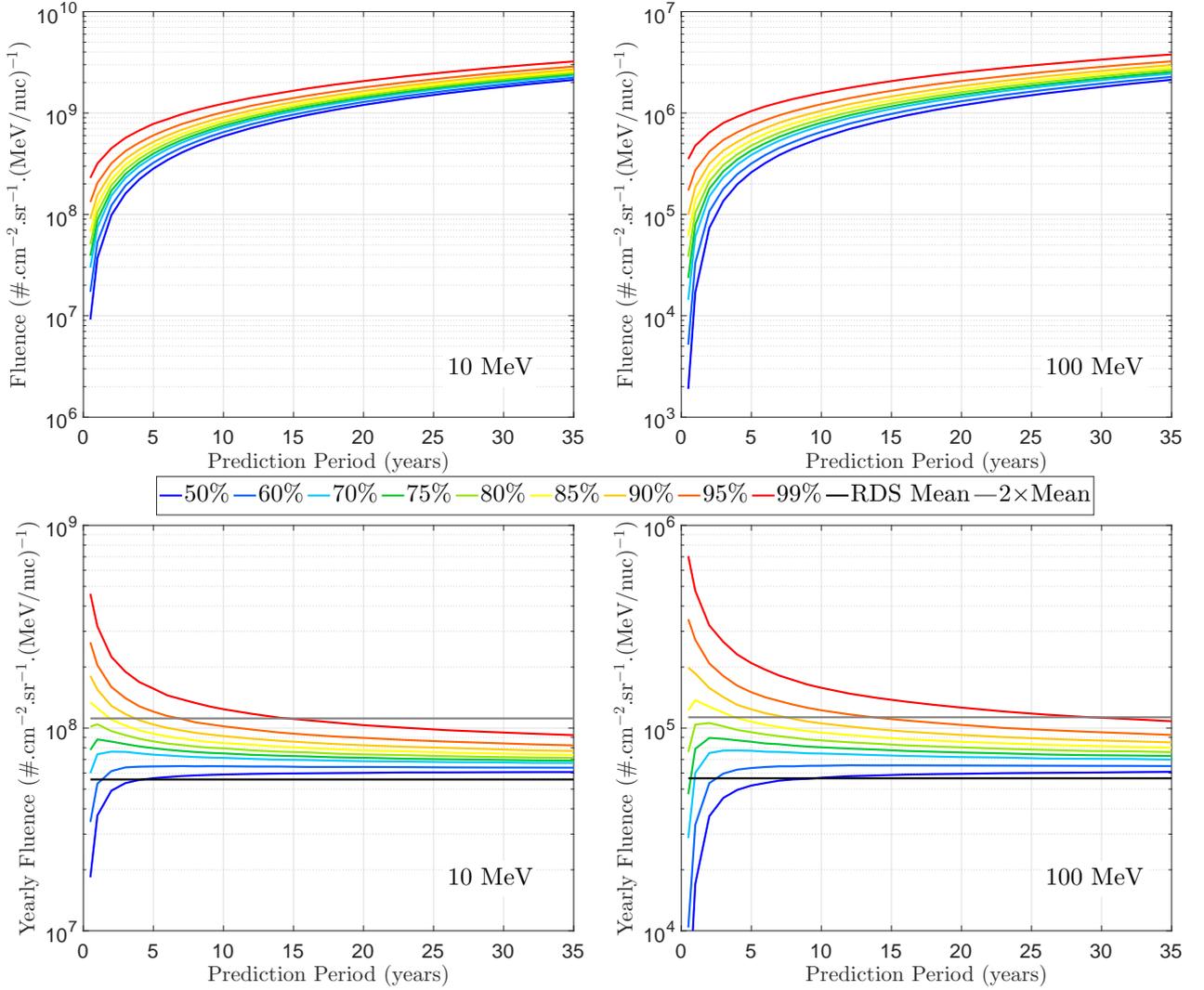}
\caption{SAPPHIRE cumulative fluence outputs at solar maximum as a function of prediction period for 10 MeV (top left) and 100 MeV (top right) and the same outputs expressed in terms of the mean yearly fluence for 10 MeV (bottom left) and 100 MeV (bottom right).}
\label{fig_flux_dur}
\end{figure}

It is interesting to look at the progression of SEP cumulative fluence outputs with prediction period and confidence. Figure \ref{fig_flux_dur} shows the model outputs at energies of 10 MeV (top left) and 100 (top right) MeV as a function of prediction period for a range of confidence intervals for solar maximum conditions. From these plots it is difficult to discern the quantitative impact of increasing the prediction period in terms of the additional fluence. The plots below show the progression of yearly fluence as a function of prediction period calculated by dividing the fluences by the model duration in each case. This shows that for lower confidence levels the yearly fluence increases with time but that for high confidence levels the yearly fluence reduces as the model regresses towards the population mean. The mean value of the population appears to be equivalent to a 60-70\% confidence level output which are flat for prediction periods above 5 years. In both cases this is significantly higher that the sample mean from the RDS indicating that the distribution fits to the data anticipate a long tail to the distribution with possible SPEs of much higher fluxes than have been observed in the space age acting to shift the population mean upward. This is a result which could not be reproduced by a data-driven model which would inevitably converge to the mean of the underlying data sample. It is also interesting to see that even for very high confidences such at 99\% that the cumulative fluence model output reduces to less than a factor of 2 of the RDS. However, this progression is much quicker for lower energies such at 10 MeV (15 years) compared to 100 MeV (30 years).

In order to explain these differences it is necessary to look at the physical processes creating SPEs. The ability of shocks to accelerate particles to very high energies is dependent on its strength. In addition to shock speed the other important factor in determining shock strength is the variability of the conditions in the ambient medium which is much higher in the corona than in the solar wind. \citet{Sandroos2009} found that shocks capable of accelerating particles to GLE levels are more easily created near regions where the coronal magnetic field is curved, e.g. in the vicinity of active regions or helmet streamers. This variability is present only in the inner corona (below $\sim3$ solar radii). In addition, there is strong lateral variability in seed particle densities in the corona which have been shown to be important factor controlling the highest energies obtained from the shock acceleration \citep{Vainio2017}. These factors give rise to very localised regions and short time durations where conditions for acceleration of particles up to 1 GeV are favourable resulting in rarer detection of higher energy particles due to the reduced likelihood of magnetic connectivity giving rise to higher intrinsic variability of higher energy particles seen at 1 AU. This explains the slower tapering of environment variability at higher energies compared to lower energies as illustrated in Figure \ref{fig_flux_dur}. This in turn indicates that the small sample of 4 solar cycles of data included in the RDS on which SAPPHIRE is based is less of an issue for the lower energies than the higher energies where there is intrinsically more variability. Future work will study the statistical errors in SEP model predictions.

\section{Conclusions}\label{sec_conc}

The proton component of the new Solar Accumulated and Peak Proton and Heavy Ion Radiation Environment (SAPPHIRE) model has been presented. The model includes results for solar minimum as well as solar maximum, outputs of cumulative fluence, large SPE fluence and peak flux for particles energies ranging from 0.1 MeV to 1 GeV. The cumulative fluence model represents a reduction in estimated fluence by as much as an order of magnitude compared to the ESP-PSYCHIC and JPL models at energies of 100 MeV but more modest reductions of up to a factor of $2-3$ at $35$ MeV more relevant for protons capable of penetrating nominal spacecraft shielding and smaller differences below 10 MeV for energies relevant for solar cell degradation effects. The SAPPHIRE proton cumulative fluence model has been compared to another model by \citet{Raukunen2017} based on data from GLEs and shows excellent agreement from 10 to 100 MeV. For higher energies, which would be important for heavily shielded environments such as the International Space Station (ISS) and for future manned missions, the difference between the models is approximately a factor of 2. Further work shall be undertaken to investigate these differences in this context in future ESA activities. The low energy extrapolation of the SAPPHIRE model has been validated against data from ACE/EPAM with excellent agreement for cumulative fluences supporting use of SAPPHIRE for applications such as solar cell degradation estimation and effects on thin films and coatings.

Comparisons of the highest SPE fluence and peak flux outputs of the SAPPHIRE proton model with the ESP-PSYCHIC models show better agreement than those for the cumulative fluence outputs although differences in the underlying data sets still result in reductions of a factor of 2 or more at 100 MeV. The flux distribution in SAPPHIRE implicitly allows for higher output spectra for high confidence, high prediction period runs with respect to ESP-PSYCHIC which is demonstrated when methods are applied to the same data set. SAPPHIRE also includes the facility to produce SPE spectra with a given mean return frequency (or a 1-in-x-year SPE). Comparisons of these outputs with ESP `Design Limit' and CREME96 spectra indicate the the CREME96 output might most appropriately be replaced by a 1-in-100-year SPE. Differences are well explained by updated data processing \citep{Sandberg2014,Rodriguez2017} and the SAPPHIRE statistical approach creating envelope output spectra contrasting with the CREME96 approach basing outputs on the fluxes from a single SPE. Comparison with ACE/EPAM data shows differences with SAPPHIRE model proton peak flux outputs at lower energies outside the core model energy range (i.e. $<5$ MeV) where outputs are extrapolated indicating that that further validation is required. It should be noted that while the lower energy range is important for dose calculations there are few effects which rely on the peak flux outputs at these energies.

The SAPPHIRE model also includes outputs for solar helium and abundances to extrapolate these results to heavier ions important for calculations of Single Event Effects (SEEs). This work is the topic of a separate publication \citep{Jiggens2018} which also includes results expressed in terms of effects quantities such as total ionising and non-ionising dose and information on how SAPPHIRE is implemented for missions mixing periods of solar maximum and minimum. It is proposed to replace existing standards in ECSS relating to SEPs with the SAPPHIRE model.

\bibliography{bibrefth}
 

\end{document}